\documentclass[11pt,preprint]{aastex}

\shorttitle{Dynamical Masses}
\shortauthors{Hillenbrand et al.}

\begin{document}


\title{
  An Assessment of Dynamical Mass Constraints on Pre-Main Sequence
  Evolutionary Tracks
}


\author{Lynne A. Hillenbrand and Russel J. White}
\affil{California Institute of Technology, Dept. of Astronomy/Astrophysics,\\ Mail Code 105-24, Pasadena, CA 91125, USA}




\begin{abstract}

We have assembled a database of stars having both masses determined from
measured orbital dynamics and sufficient spectral and photometric
information for their placement on a theoretical HR diagram.  Our sample
consists of 115 low mass ($M < 2.0$ M$_\odot$) stars, 27
pre-main sequence and 88 main sequence.  We use a variety of available
pre-main sequence evolutionary calculations to test the consistency of
predicted stellar masses with dynamically determined masses.
Despite substantial improvements in model physics over the past decade,
large systematic discrepancies still exist between empirical and
theoretically derived
masses.  For main-sequence stars, all models considered predict masses
consistent with dynamical values above 1.2 M$_\odot$, some models predict
consistent masses at solar or slightly lower masses, and no models predict
consistent masses below 0.5 M$_\odot$ but rather all models systematically 
under-predict such low masses by 5-20\%. The failure at low masses stems 
from the poor match of most models
to the empirical main-sequence below temperatures of 3800 K where 
molecules become the dominant source of opacity and convection is the
dominant mode of energy transport. For the pre-main sequence sample we find
similar trends.  There is generally good agreement between
predicted and dynamical masses above 1.2 M$_\odot$ for all models.  Below
1.2 M$_\odot$ and down to 0.3 M$_\odot$ (the lowest mass testable) 
most evolutionary models systematically under-predict the 
dynamically determined masses by 10-30\% on average with the Lyon group models
(e.g. Baraffe et al. 1998) predicting marginally consistent masses 
{\it in the mean} though with large scatter.  
Over all mass ranges, the usefulness of dynamical
mass constraints for pre-main sequence stars is in many cases limited by
the {\it random} errors caused by poorly determined luminosities and
especially temperatures of young stars.
Adopting a warmer-than-dwarf 
temperature scale would help reconcile the {\it systematic} pre-main sequence
offset at the lowest masses, but the case for this is not compelling
given the similar warm offset at older ages between most sets of
tracks and the empirical main sequence.  Over all age ranges, the systematic 
discrepancies between track-predicted and dynamically determined masses appear 
to be dominated by inaccuracies in the treatment of convection and in
the adopted opacities.  


\end{abstract}




\section{Introduction}\label{sec_intro}

Three of the most fundamental stellar parameters are mass, angular
momentum, and composition, which together determine almost exclusively the
entire evolutionary history of any given (single) star.  Although stars
spend the vast majority of their lives on the main sequence of
hydrogen-burning, particularly interesting stellar objects are often those
in the shorter-lived 
pre-main sequence or post-main sequence evolutionary phases. Our focus
here is on the inference of stellar masses for pre-main sequence and young
main sequence objects, for which observational data relevant to their
location in the Hertzsprung-Russell (HR) diagram has become abundant in
recent years.  Masses and ages are often inferred from such HR diagrams via
comparisons to an increasingly large suite of pre-main sequence
evolutionary calculations.  Instead of adopting a main sequence
mass-luminosity relationship, one explicitly accounts for the evolution of
the mass-luminosity relationship with age. The inferred stellar masses and
ages are then used to construct initial mass functions and to surmise star
formation histories of molecular clouds.  

The pre-main sequence luminosity and effective temperature evolution of
just-born stars was first calculated over a range of masses by Iben (1965)
and by Ezer and Cameron (1967a,b) who assumed homologous contraction and
solved the equations of stellar structure following the formalism of
pioneers Henyey and Hayashi.  Substantial improvements in the input physics
and opacities was achieved during the following decades by several others,
notably VandenBerg (1983) and D'Antona \& Mazzitelli (1985). In the 1990's
several series of papers by different groups incorporated yet more complex
and varied assumptions regarding the equation of state, opacities,
convection physics, outer boundary condition of the stellar interior, and
treatment of atmospheres.  Electronically available data from these
calculations including those from Swenson et al. 1994 (S93); D'Antona \&
Mazzitelli 1994 (DM94) and 1997 (DM97); Forestini 1994, Siess et al. 1997,
and Siess et al. 2000 (S00); Baraffe et al. 1998 (B98) and Chabrier et
al. 2000; Palla \& Stahler 1993 and 1999 (PS99); and finally Yi et al. 2003
(Y$^2$) were widely circulated. Other authors such as Burrows et al. 1997
and Baraffe et al. 2002 have focussed on sub-stellar mass objects.  

Complications to simple luminosity and effective temperature evolution
via radial contraction are the effects of rotation, composition, accretion,
magnetic fields, and the presence of dust in the atmospheres of the lowest
mass stars and brown dwarfs.  These have been explored in limited capacity
as well, as discussed by Mendes et al. (1999), D'Antona et al. (2000),
Baraffe et al. (2002), and Siess et al. (1997a,b).  In
addition, the ``zero point" or initial mass-radius relationship from which
pre-main sequence evolution begins is poorly constrained (see Larson, 1972;
Stahler 1983; Mercer-Smith, Cameron, \& Epstein 1984; Palla \& Stahler,
1993; Bernasconi 1996; Hartmann, Cassen, Kenyon 1997; Baraffe et al. 2002).
In addition, comparison between data young cluster data and isochrones,
including the predictions from lithium burning, show inconsistencies which
lead us to infer that
ages younger than $\sim$10 Myr are particularly uncertain, 
and masses are also likely biased.  Despite the large uncertainties
and indeed the cautions offered by many of the above authors themselves
regarding the utility of their models in explaining observations, the
existing array of models has been used heavily over the past decade for
comparison to the H-R diagrams assembled for pre-main sequence stars in
nearby star-forming regions.  These tracks are the primary tool used to
determine the ages and masses of young stars, and thus a cornerstone on
which the conclusions of many star formation studies rest.  Two examples are
the interpretation of observational data in a meta-context such as 
the initial mass function or the star formation history of a particular
region, or the evolution of circumstellar disks or stellar angular momentum
through the pre-main sequence.  Such conclusions rely entirely on the 
evolutionary models and systematically different results can arise from the
use of different models.

Fundamental calibration of pre-main sequence evolutionary tracks is,
however, not yet established.  Several tests have been proposed.  The
predicted masses can be compared to those inferred from either binary 
orbits (e.g. Casey et al. 1998; Covino et al. 2000; Steffen et al. 2001) or
velocity profiles of rotating circumstellar disks (e.g. Simon et al. 2000;
Dutrey et al 2003).  The predicted ages can be compared under the
assumption of co-eval formation, to loci of pre-main sequence binaries
(e.g. Hartigan, Strom, \& Strom 1994; Prato, Greene, \& Simon 2003), higher
order multiples (White et al. 1999) and young ``star-forming" clusters
(e.g. Luhman et al. 2003; Hillenbrand, Meyer, \& Carpenter 2004).  Older
open clusters offer even narrower sequences for comparison with model
isochrones (e.g. Stauffer, Hartmann \& Barrado y Navascues 1995).  All of
these tests, however, are limited by the accuracy with which individual
stars can be placed on a theoretical H-R diagram.  In addition to the
poorly understood observational errors, uncertainties in the temperature
and bolometric correction scales themselves remain significant, especially
at sub-solar masses and young ages.

In this paper we explore the consistency of the masses predicted by various 
sets of pre-main sequence evolutionary tracks with those masses 
fundamentally determined from orbital dynamics.  
Our sample is larger than those considered in
previous experiments (referenced above); in particular we include both
pre-main sequence and main sequence stars.  
The lower mass limit in our sample is imposed by the available
fundamental mass data (0.1 M$_\odot$ for main sequence stars but only
0.3 M$_\odot$ for pre-main sequence stars) 
and the upper limit (2.0 M$_\odot$) is adopted to include only
un-evolved main sequence objects.

In section ~\ref{sec_tracks} we discuss the models we test and the systematic
differences between them.  Section
~\ref{sec_data} presents the database of double-lined binaries or
single/multiple stars harboring rotating gaseous disks with determined stellar
masses, and our methodology for inferring masses from pre-main sequence
evolutionary calculations.  In section ~\ref{sec_comp} we perform the
detailed comparison of the model masses and the dynamically-determined 
fundamental masses.  Section ~\ref{sec_conc} contains our conclusions
and recommendations.

\section{Pre-Main Sequence Evolutionary Models}\label{sec_tracks}

The various sets of tracks available and their most basic input assumptions
regarding stellar interior structure and physics are reviewed in this
section.  
In our analysis we make use of those sets of models which have been 
made available electronically by the authors.  We refer the interested reader 
to the references cited for more detail on individual sets of calculations. 
We do not attempt to assess the physical validity, triumphs, or shortcomings 
of the individual models; we present them purely for consideration in 
comparison to stellar masses fundamentally determined based on astrophysical 
data.

\subsection{Victoria Group: S93}

The heritage of the Swenson et al. models resides in the 
Victoria stellar evolutionary code of VandenBerg (1983; 1992).
The notation ``S93" refers to a private communication in 1993 
of approximately the series F models described in Swenson et al. 1994, 
provided initially to K. Strom and subsequently to the 
present authors.  The mass range covered is 0.15-5.00 M$_\odot$.
These models employ the OPAL (Rogers \& Iglesias 1992) and Cox \& Tabor (1976)
opacities, an ``improved" 
Eggleton, Faulkner, \& Flannery (1973) equation of state,
Fowler et al. (1975) and Caughlan \& Fowler (1988) reaction rates,
use a mixing length parameter $\alpha =$ 1.957, and assume abundances
Y = 0.282 and Z = 0.019. 
Their starting point is defined as $\rho <$ 0.01 g/cm$^3$.
Atmospheric treatment is presumed grey.
A hint provided in VandenBerg \& Clem (2003) suggests that more may be coming
from this group on pre-main sequence evolution including realistic
atmospheres, with the most recent
description of main sequence and post-main sequence evolution appearing
in VandenBerg et al. (2000).

\subsection{D'Antona and Mazzitelli: DM94 and DM97}
D'Antona \& Mazzitelli (1994) provided tracks covering is 0.1-2.5 M$_\odot$
using the first substantial improvement to input physics since 
the 1980's pre-main sequence evolutionary papers which utilized 1970's era 
physics.  The models employ the Alexander et al. (1989) or Kurucz (1991) and
Rogers \& Iglesias (1992) opacities, Mihalas et al. (1988) and Magni \& Mazzitelli
(1979) equation of state, Caughlan \& Fowler (1988) 
and Fowler et al. (1975) reaction rates,
use either a mixing length parameter $\alpha =$ 1.2 or the newly introduced
Canuto \& Mazzitelli (1991, 1992) ``full spectrum of
turbulence" (FST) convection prescription as a rival to the standard
mixing length theory (MLT), and assume abundances Y = 0.285 and Z = 0.018. 
Atmospheric treatment is grey.
Their starting point is the sequence of deuterium burning.
These models were updated to cover 0.017-3.00 M$_\odot$
in D'Antona \& Mazzitelli 1997 (DM97) and again in 1998
(the later being a ``web-only" correction at $<$0.2 M$_\odot$
to the originally circulated 1997 models).  As this article went to press
we became aware of the Montalban et al. (2004) calculations which explore
both MLT and FST convection and now use the non-grey 
Hauschildt, Allard, \& Baron (1999) or the Heiter et al (2002) a.k.a. Kurucz
atmospheres.  These models are not electronically available at present
and are not used in our analysis.

\subsection{Geneva group:  C99}

The Charbonnel et al. 1999 models cover 0.4-1.0 M$_\odot$
and represent an extension to lower masses of the Geneva code.  
They employ the MHD 
(Hummer \& Mihalas 1988; Mihalas et al. 1988; Dappen et al 1988)
equation of state, the Alexander \& Fergusson (1994) and 
Iglesias \& Rogers (1996) opacities,
Caughlan \& Fowler (1988) reaction rates,
a mixing length parameter $\alpha =$ 1.6, 
and abundances Y = 0.280 and Z = 0.020. 
The atmospheric treatment down to $\tau=2/3$ is grey.
These models are not publicly available and are not utilized in 
the present study.

\subsection{Palla and Stahler: PS99}

The Palla \& Stahler (1999) models over 0.1-6.0 M$_\odot$
use the Rosseland mean opacity,
the Eggleton, Faulkner, \& Flannery (1973) and Pols et al (1995)
equation of state, Fowler et al. (1975) and Harris et al. (1983) reaction rates,
a mixing length parameter $\alpha =$ 1.5, 
and assume abundances Y = 0.28 and Z = 0.02. 
The calculations explicitly include a
``birthline" or initial mass-radius relationship (which, incidently, could be
adopted and independently applied by any of the other calculations reviewed
in this section).  Atmospheric treatment is grey.
These models do not extend beyond ages of 10$^8$ years.

\subsection{Grenoble group:  S00}

The Grenoble group has published their calculations in
Forestini 1994, Siess et al. 1997, and 
most recently in Siess et al. 2000 (S00). 
The calculations cover 0.1-7.0  M$_\odot$.  They
use the Alexander \& Fergusson (1994) and 
Iglesias \& Rogers (1996) opacities,
a modified Pols et al. (1995) scheme for the equation of state,
Caughlan \& Fowler (1988) reaction rates,
a mixing length parameter $\alpha =$ 1.6, 
and abundances Y = 0.288 and Z = 0.0189. 
These models attempt to include a ``realistic" 
atmosphere as the outer boundary condition using data from Plez (1992) and
Kurucz (1991).

\subsection{Lyon group: B98}

The Lyon group published models in Baraffe et al. 1995,
Chabrier \& Baraffe 1997, and Baraffe et al. 1998 (B98). 
The calculations cover 0.035-1.2 M$_\odot$; 
see Chabrier et al. 2000 and Baraffe et al. 2002 
for an extension to 0.001 M$_\odot$.  The Lyon group
uses the Alexander \& Fergusson 1994 and Iglesias \& Rogers 1996 opacities,
the Saumon, Chabrier, \& van Horn 1995 equation of state,
reaction rates described in Chabrier \& Baraffe 1997,
several different values for the mixing length parameter 
$\alpha =$ 1.0, 1.5, and 1.9, and 
abundances Y = 0.275, 0.282 and Z = 0.02. These models also employ the 
non-grey Hauschildt, Allard, \& Baron (1999) atmospheres which include 
molecular opacity sources such as TiO and H2O, as well as dust grains. It
should be noted that the $\alpha =$ 1.9 models are actually the same as  
the $\alpha =$ 1.0 ones below 0.6 M$_\odot$, and also that the
$\alpha =$ 1.9 models actually use $\alpha =$ 1.0 in the atmospheres,
at optical depths $<$100.
The B98 models do not extend to radii larger than those defined by
the 10$^6$ year isochrone limiting their utility in studies of young
low-mass star-forming regions where populations are frequently found above
the limit of the B98 tracks.

\subsection{Yale group: Y$^2$ and YREC}

The Yale group has two current sets of models, one
called  ``Y$^2$" and the other ``YREC" which includes rotation.

The Y$^2$ models cover 0.4-5.0 M$_\odot$ and have been published 
in a series of papers:
Yi, Kim, \& Demarque 2003; Kim et al. 2002; Yi et al 2001.
These models use Iglesias \& Rogers 1996 
and Alexander \& Fergusson 1994 opacities,
the Cox \& Giuli 1968 and 
Rogers et al. 1996 equation of state with implementation of the
Debye-Huckel correction (Guenther et al. 1992), 
reaction rates from Bahcall \& Pinsonneault 1992,
a mixing length parameter $\alpha =$ 1.7431, and a range of abundances 
where we have chosen the X = 0.71, Y = 0.27,  Z= 0.02 models 
for comparison.  Atmospheres are presumed grey but for the purpose of 
calculating colors (not relevant to the present study) are matched in a 
semi-empirical way to the
color-temperature relations adopted by Lejeune et al. (1998)
\footnote{We note specifically that the V-K vs log T$_{eff}$ relationship
given in Lejeune et al. does not place the M dwarfs in the present study on 
the main sequence, but rather substantially warmer and fainter than 
the main sequence.}. These models begin at the theoretically defined
deuterium burning main sequence. 

The YREC (Yale Rotating Evolution Code) models
cover 0.1-2.25 M$_\odot$ and have been published 
in Guenther et al 1992 and Sills, Pinsonneault, \& Terndrup 2000.
Currently these models also use Iglesias \& Rogers 1996 
and Alexander \& Fergusson 1994 opacities,
the Rogers et al. 1996 but also the Saumon, Chabrier \& Van Horn 1995 
equations of state as appropriate, 
reaction rates from Gruzinov \& Bahcall 1998,
a mixing length parameter $\alpha =$ 1.72, and abundances 
corresponding to Y=0.273 and Z= 0.0176 at the age of the Sun.
The atmospheric treatment is the same as Y$^2$.
These models are not publicly available and are not utilized in 
the present study.  

\subsection{Comparison of Models and Systematic Effects}

\begin{figure}
\epsscale{1.0}
\plotone{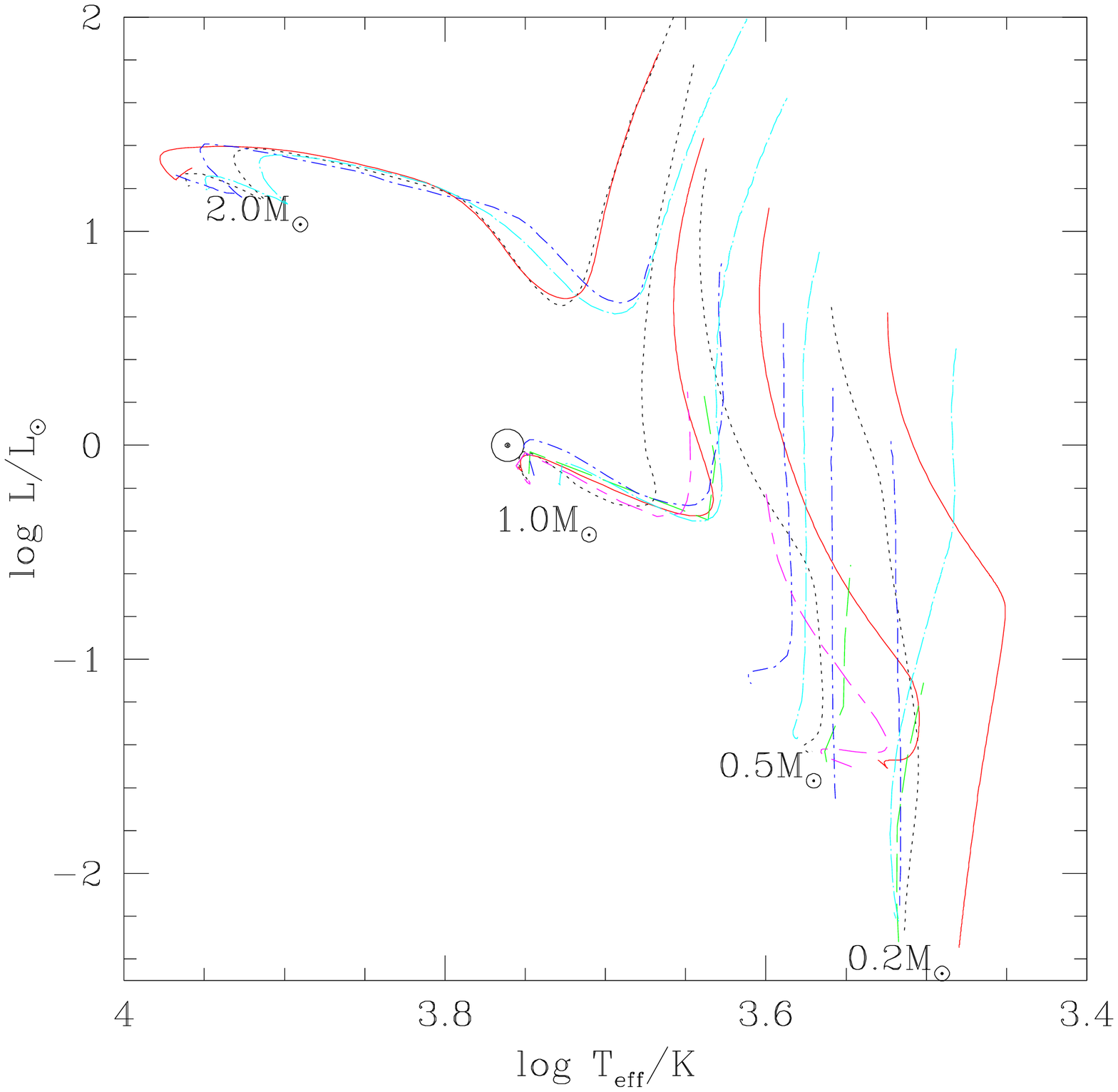}
  \caption{
  Variation between pre-main sequence contraction tracks
  for masses 0.2, 0.5, 1.0, and 2.0 M$_\odot$.  Line types indicate models of 
  Swenson et al. (1993 -- solid line), 
  D'Antona \& Mazzitelli (1997/1998 - dotted line), 
  Baraffe et al. $\alpha$=1.9 (1998 -- long-dash line), 
  Palla \& Stahler (1999 -- dot-short-dash line), 
  Siess et al. (2000 --dot-long-dash line), and
  Yi et al. (2003 -- long-dash-short-dash line). 
  Note that the Palla \& Stahler models, 
  for which no 0.5 M$_\odot$ track is available,
  have both the 0.4 and the 0.6 M$_\odot$ tracks plotted instead.  Also note
  that the Yi et al. models do not extend as low as 0.2 M$_\odot$.
  \label{comparetrack} }
\end{figure}

\begin{figure}
\epsscale{1.0}
\plotone{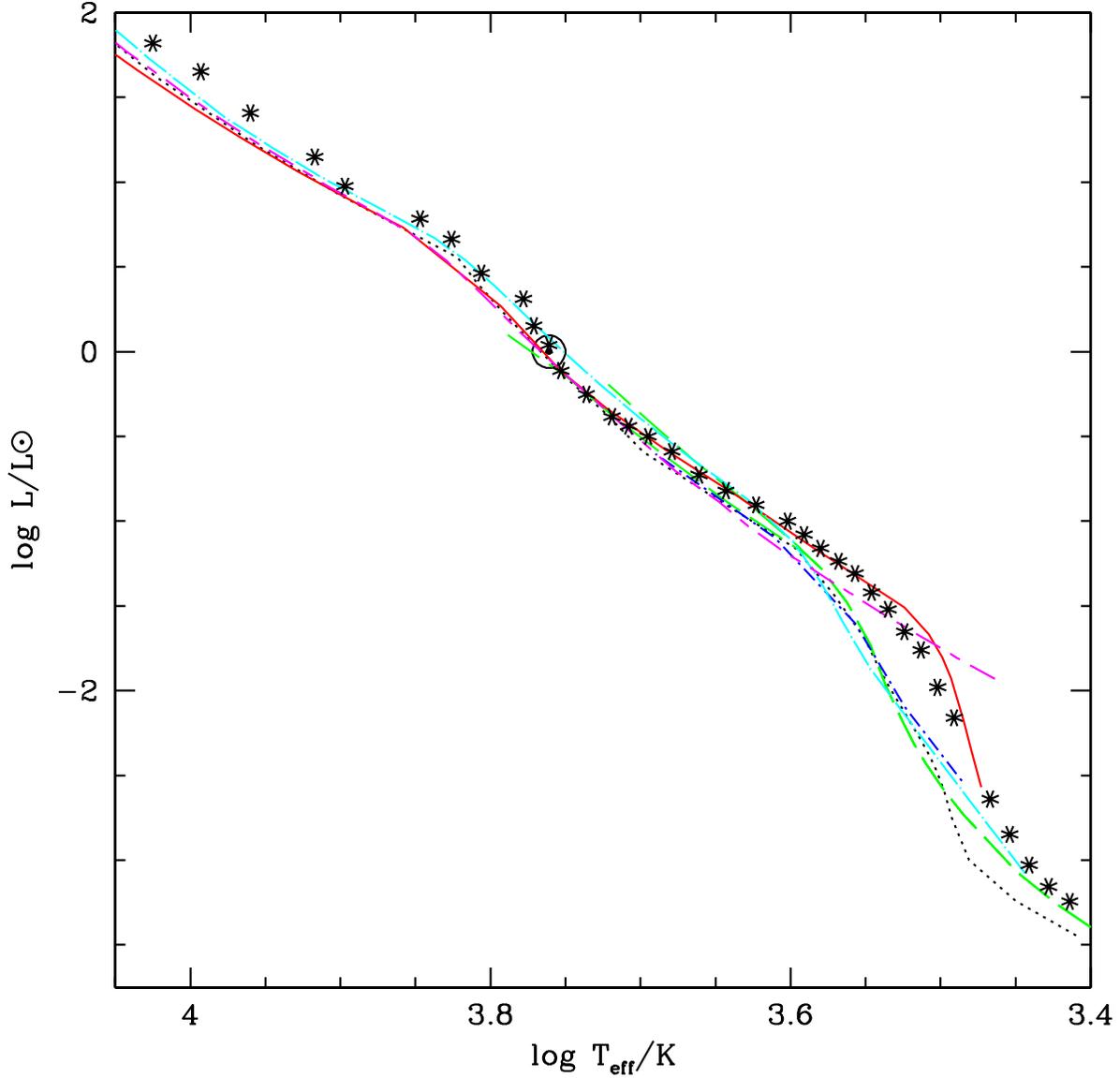}
  \caption{
  Comparison of the composite main sequences adopted by us using
  the various evolutionary models.  Line types are the same as in  
  Figure~\ref{comparetrack}. Asterisks show the ``empirical"
  main sequence derived from measurements of M$_V$ and our
  adopted dwarf bolometric correction and temperature scales (see Appendix).  
  Note that the empirical main sequence represents the average observed
  luminosity as a function of temperature along the main sequence and not
  necessarily the zero-age main sequence.  Consequently, the highest mass
  main sequence stars are on average more evolved relative to zero-age than
  the average solar-mass main sequence star; this likely causes the
  apparent over-luminous location of the empirical main-sequence at higher
  masses.  The Swenson et al. and the Yi et al. models extend cool enough
  to more accurately reproduce the low mass empirical main sequence 
  compared to the other calculations; note however the ``straight"
  nature of the Yi et al. main sequence which is at odds with the
  expected downturn due to H$_2$ dissociation.
  \label{zamshrd} }
\end{figure}

As illustrated by the above discussion of the gamut of pre-main sequence
evolutionary models, there is substantial variation in the treatment of
various aspects of the physics as well as in the adopted values of certain
parameters.  The most salient of these differences are in the opacity
sources, treatment of convection, and treatment of interior/atmospheric
boundary conditions.  For comparison between the results of several of the
above-mentioned codes at low masses, we show in Figure \ref{comparetrack}
the predicted contraction of different stellar masses, and in Figure
\ref{zamshrd} the resulting zero-age main sequence as defined in Section
3.4. Systematic differences are apparent in the mass tracks, especially at
young ages, and on the main sequence, particularly at low masses.  
The variations between tracks are predominantly in temperature and 
only secondarily in luminosity.


The predicted effective temperature for a given mass star is dictated
largely by the treatment of convection in both the atmosphere and the
interior.  Because of the extreme complexity of a realistic prescription,
convection is usually handled by adopting the mathematically simple
mixing-lenth theory (B\"ohm-Vitense 1958), although more sophisticated
prescriptions have been proposed (e.g. Canuto \& Mazzitelli 1992). 
Typically, larger mixing lengths (more efficient convection) predict hotter
evolutionary tracks and yield lower masses for a given position in the
HR diagram.  The choice of the mixing
length is a large uncertainty in current models.  A common value is one
which predicts 1 M$_\odot$ model agreement with the solar model, but this 
approach may artificially compensate for other inadequacies in the 
calculations.  For example,
several other major aspects of convection can affect the track temperatures
such as how the interior is matched to atmosphere, the thickness of the
convective region, and the extent of convective overshooting (see
e.g. D'Antona \& Mazzitelli 1994; Montalban et al. 2004). Consequently, the
treatment of convection is one of the primary uncertainties in current
evolutionary models.  A related effect is the opacity (including the
influence of metallicity) through which the convective energy transport must
occur.  Higher opacities generally mean lower predicted effective
temperatures for a given mass star.

Another point of comparision between sets of models is the match between
the various 1 M$_\odot$ tracks and the location of the Sun. The Sun is
evolved from its zero-age main sequence location being hotter, larger, and
more luminous. In some cases certain parameters in the above sets of models
have been adjusted by the model authors such that their 1 M$_\odot$ model 
reproduces the temperature and luminosity of the present day Sun.  This requires
that the model tracks extend beyond the zero-age main sequence.  Nevertheless, 
we illustrate in Figures \ref{comparetrack} and \ref{zamshrd} the 
location of the Sun compared to 1 M$_\odot$ pre-main sequence tracks and 
zero-age main sequences (effectively the 10$^8$ year isochrone at this mass;
see \S 3.4) from 
various models.  This comparison notwithstanding, we demonstrate in our results 
that there is little correspondence between models being able to match 
the observed main sequence parameters and the observed pre-main sequence parameters.

Finally, it should be stressed that there is generally poor agreement between
the various models and the empirical main sequence at low masses
(Figure \ref{zamshrd}).
Of note is that the Y$^2$ models which at low masses do seem to reach 
as cool as the empirical data, do not display the same downturn at low
temperatures as other models.  A downturn, such as that displayed by the S93
models in the same cool regime, and by the other models at much warmer
temperatures, is expected based on the dissociation of H$_2$ 
(Copeland et al 1970).

\section{Astrophysical Data}\label{sec_data}

\subsection{Sample and Selection Criteria}

In order to test the predictions of the various pre-main sequence
evolutionary tracks just discussed, we have compiled from the literature a
list of stars with dynamically determined masses and with luminosity and
temperature estimates for placing them on the HR diagram.  The sample is
restricted to stars less massive than 2.0 M$_\odot$.  Of the 148 stars in
this sample (Table 1), 88 are main-sequence and 27 are pre-main sequence stars;
the remaining 33 stars are determined to be post-main sequence as 
described below.  The Sun is included as a main sequence star with stellar 
parameters adopted from Gray (1992).

For the main sequence sample we require masses measured to better than
10\%.  We strive to exclude W UMa type contact binaries (e.g. V781 Tau; 
Liu \& Yang 2000) in which tidal effects or mass transfer could be important. 
Further, to avoid including stars evolved too far beyond the
zero-age main-sequence (ZAMS), we have retained for analysis only those
binary components in Table 1 with log g $>$ 4.20 cm sec$^{-2}$, and thus stars 
less evolved than $\sim$600 Myr from the ZAMS near our upper mass range and 
less evolved than 1-3 Gyr from the ZAMS near the solar mass range (according 
to the Girardi et al. 2000 post-ZAMS models).  
We begin with the catalog of Anderson (1991) and the
additional lists compiled by Ribas (2000), Delfosse et al. (2000), and
Lastennet \& Valls-Gabaud (2002), but also include systems more recently 
identified in Munari et al. (2001), Zwitter et al. (2003),
and Marrese et al (2003).  
Of the compiled systems surviving our selection criteria, 
most are detached double-lined
eclipsing binaries. The remaining main-sequence stars are spatially
resolved double-lined spectroscopic binaries which have independent
temperature estimates for each component from spectroscopic or color
measurements that enable their placement on the HR diagram.  We note that the
main-sequence sample of stars suitable for our purposes has, historically,
been biased towards solar or greater masses.  In recent years, however,
the sample of stars at masses $<$0.5 M$_\odot$ with both dynamical masses 
and independent temperature and luminosity estimates for the two components 
has grown considerably (e.g. Delfosse et al 2000).


The pre-main sequence sample is not subjected to the same dynamical mass
uncertainty restriction that is applied to the main sequence sample
($\sigma< 10$\%) due to the small numbers of stars having 
measured masses.  These 27 pre-main sequence stars include 8 components of
double-lined eclipsing binary systems (TY Cr Ab, EK Cep B, RS Cha A \& B,
RXJ0529.4+0041 A \& B, AK Sco A \& B; see references in Table 1) which have 
the most accurately determined masses among the pre-main sequence sample 
($\sigma \le 5$\%), but are all approximately solar or larger mass stars.  One
pre-main sequence system has component masses determined from spatially
resolved measurements of a double-lined spectroscopic binary (NTTS
045251+30016 A \& B; Steffan et al. 2001).  Nine pre-main sequence stars
have masses determined from disk kinematics (Simon et al. 2000; Dutrey et al. 2003). 
In the
case of the UZ Tau E binary, the component masses are determined from the
spectroscopic orbit inferred by Prato et al. (2002).  The remaining
pre-main sequence systems (FO Tau, FS Tau, DF Tau, GG Tau) are all binaries
which have only total dynamical mass estimates; in these cases we thus
compare these total dynamical masses to the summed masses inferred from
placement of the individual components on the HR diagram.  Although other
pre-main sequence binary systems have orbital mass estimates, we include
only those which have spatially resolved temperature or spectral type 
measurements.  We do not include systems with only mass ratios available.

\subsection{Stellar Parameters I: Mass, Radius and Surface Gravity}

\begin{figure}
\epsscale{1.0}
\plotone{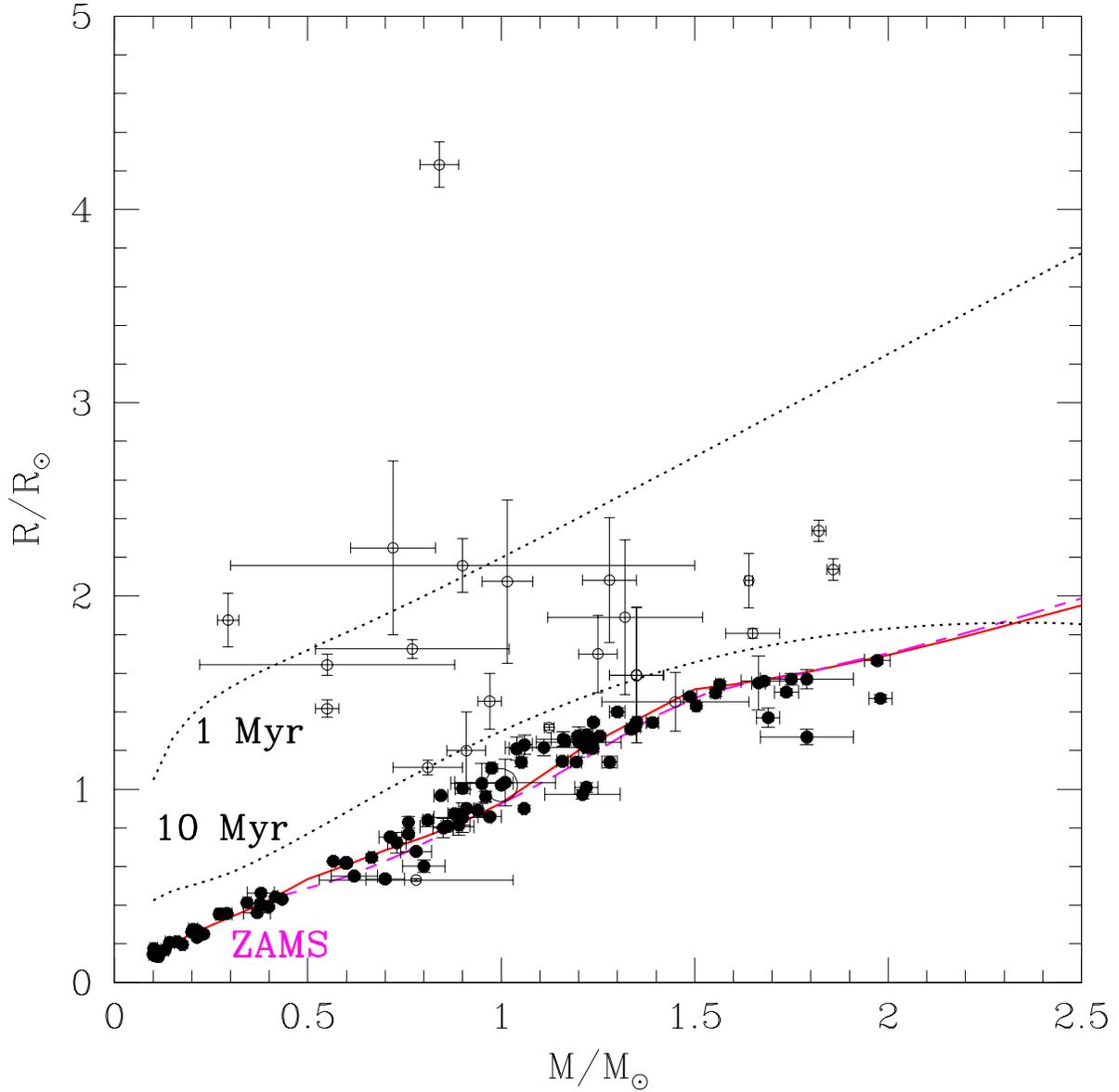}
  \caption{
  Mass and radius measurements for our sample stars.  Open symbols represent 
  pre-main sequence objects while filled symbols are main sequence stars.
  For  
  the double-lined eclipsing systems both axes are fundamentally derived from 
  observation whereas for the non-eclipsing systems 
  the masses are fundamental but the radii are inferred
  from luminosity and effective temperature values in Table 1.
  The 1 and 10 Myr isochrones of D'Antona \& Mazzitelli are indicated 
  (dotted lines) to show
  the approximate change in radius with age as pre-main stars contract, as are
  the ZAMS from S93 (solid line) and Y$^2$ (dashed line) which most closely
  approximate the empirical main sequence in Figure \ref{zamshrd}.
  \label{massrad} }
\end{figure}

The sample is listed in Table 1 in order of the most to the least massive
star and with pre-main sequence stars distinguished from main sequence
stars.  The mass and radius range occupied by the un-evolved members of
our sample (log g $>$ 4.20 cm sec$^{-2}$) is shown in Figure
\ref{massrad}.  For stars which are members of eclipsing systems, radii
are determined directly from observations; for the remainder this quantity
has been estimated for plotting purposes from temperature and luminosity
following Stefan's Law (L = 4$\pi$R$^2\sigma T_{eff}^4$).  In the remainder of
this section we describe how the masses, radii, and gravities listed 
in Table 1 were derived  by the original authors.

For the double-lined eclipsing binaries, the ratio of velocity amplitudes
is inversely proportional to the ratio of masses while the sum of velocity
amplitudes is related via the period to the sum of the masses.  Given two
equations and two unknowns, the individual component masses can thus be
determined directly from the observables $v_1$, $v_2$, and the orbital
period.  Photometric measurements of the eclipse provide the ratio of
radius to semi-major axis while the assumption of $\approx$90 degree system
inclination means that radial velocity measurements yield the semi-major
axis uniquely and hence one can solve for the radius directly from the
observations (e.g. Covino et al. 2000).  Double-lined eclipsing binary
systems are the only binary systems with radius estimates determined
directly from observables.  The radii combined with the masses yield
surface gravities ($g$ = GM/R$^2$).  Only those radii and surface gravities
determined from fundamental observables are listed in Table 1.

For the spatially resolved double-lined systems, one does not have the
benefit of knowing the system inclination.  Instead, one can constrain the 
inclination via a combined astrometric and radial velocity orbital
solution, allowing the individual masses to be recovered (e.g. Steffen et
al. 2001).  For spatially resolved binaries with an astrometric orbital
solution but no radial velocity orbital solution, a total system mass can
be determined if a distance is assumed (e.g. Schaefer et al. 2003).
Finally, for stars surrounded by spatially resolvable circumstellar gas
disks, interferometric measurements which map the velocity profile can be
used to dynamically determine the central mass under the assumption of
keplerian motion (e.g. Simon et al. 2000).  In some cases, the central mass
may a binary star.

\subsection{Stellar Parameters II: Temperature and Luminosity}

Comparison of the dynamically determined masses discussed above with those
inferred from theoretical calculations requires temperature and luminosity
information for every star.  In determining these values, we apply the same
methods to both the main-sequence and the pre-main sequence samples.  For
the eclipsing binary systems, the ratio of the stellar temperatures are
typically determined very precisely from light curve analysis (see
individual references cited in Table 1).  These values are then combined with
a mean system temperature estimated from photometrically calibrated
atmospheric models (see e.g. Ribas et al. 2000), to detemine individual
effective temperatures\footnote{Several eclipsing systems are known to be 
chromospherically active binary stars (e.g. Strassmeier et al. 1993) in
which star spots are an unaccounted-for bias in the temperature
estimates. These systems are noted as such ("CABS") in Table 1.}. Although
the temperatures listed in Table 1 are all taken directly from the
references and thus in many cases are determined in a non-uniform fashion,
we have in all cases adopted the values which use the most recent and
accurate photometric calibrations.  Since the stellar radius is also a
quantity inferred from light curve analysis, luminosities are then 
determined directly from Stefan's law and are, for most part, distance
independent.  In some cases we transformed quoted M$_{bol}$ values to log L
values.  We assume M$_{bol(\odot)}$ = 4.75 mag in all calculations (Allen
\& Cox 2000; see also footnote \#7 in VandenBerg et al. 2000).

For the remaining (non- double-lined eclipsing) main sequence stars, we
determine temperatures in one of 3 ways.  Preferably, we adopt temperatures
and uncertainties from the listed references when they are determined from
a line ratio analysis (e.g. Gray 1994).  Alternatively, we estimate the
temperatures from the spectral types or, if no spectroscopic information is
available, we determine the temperature from the observed photometric
colors.  We use the temperature / spectral type / color
relations listed in the Appendix and assume an uncertainty in log T of 0.015
dex, which corresponds to roughly 1 spectral subclass, despite that the
formal errors in log T based on color errors would be substantially smaller.  
Temperatures
determined from either spectral types or colors are listed in italics in
Table 1 as they are not fundamental temperature measurements.  
Luminosities are recalculated here based on optical or infrared
photometry, bolometric corrections from the Appendix, and distance
estimates.  All of the main-sequence stars have parallax information, and
hence distances.  Although the luminosities are recalculated to
ensure no systematic errors from different assumptions, we generally adopt
the published luminosity uncertainties.  For the
stars with only spatially resolved photometry, we adopt
a uniform uncertainty in log L of 0.05 dex.

For the remaining (non- double-lined eclipsing) pre-main sequence stars,
temperatures are determined from spectral synthesis in the case of BP Tau
(Johns-Krull et al. 1999), or from spectral types and the temperature
relation listed in the Appendix, assuming an uncertainty of 0.015 dex.
Photometric colors alone are insufficient for estimating the temperatures
of pre-main sequence stars because of possible extinction and continuum
excesses from either an accretion shock or the inner circumstellar disk.  The 
luminosities are calculated from I$_c$-band measurements, which are the least 
likely to be contaminated by possible continuum excesses, and are at an
optimal 
wavelength from which to apply a bolometric correction for early A through
mid-M spectral types.  All pre-main sequence stars for which
we have calculated luminosities are in Taurus; we assume a distance of 140 pc 
(Kenyon et al. 1994).  Also for this sub-sample of young T Tauri
stars, we assume a uniform uncertainty of 0.10 dex in log L which incorporates
typical 1 spectral sub-class errors propogated to errors in
intrinsic colors and in bolometric corrections used
to calculate reddening-free luminosities.

\subsection{Masses Estimated from Tracks}

We derive track-predicted masses for our sample by interpolating between
tabulated luminosity and effective temperature values as a function of stellar
mass and age for each set of tracks we test.  In practice, the methods
adopted to determine masses for the main sequence and pre-main sequence
stars differ slightly.  For the pre-main sequence stars, isochrones are
generated at log ages intermediate to those tabulated by the model
authors.  The mass is determined via interpolation along the isochrone 
that intersects with the stellar luminosity and temperature.  For stars
with luminosities that put them above the youngest isochrone, the mass is
assigned using this youngest isochrone and the temperature.  This occurs only 
for the B98 tracks and only for a few late-K and M type T Tauri stars.  
Uncertainties in the
track-predicted masses are determined from the range of masses predicted by 
varying the luminosity and temperature estimates by their uncertainties
as listed in Table 1.


For stars already on the main sequence where isochrones converge in the
luminosity / effective temperature plane, we have created a theoretical
young main sequence for each set of tracks by adopting the 10$^8$ year
isochrone at masses of 0.7 M$_\odot$ and above (such that stars have
already arrived at their ZAMS position but have not yet begun any substantial
evolution away from it) and the 10$^9$ year isochrone below this mass.  
Only objects less massive than 0.09 M$_\odot$ have not reached the ZAMS by 
10$^9$ years according to the models; the least massive main sequence
star in our sample
is 0.10 M$_\odot$.  We refer to Figure \ref{zamshrd} for comparison of the
luminosity / effective temperature relationship adopted as the main
sequence for the various sets of tracks.  These constructed
main-sequences represent a unique mass-temperature and mass-luminosity
relation for each model.  We use these relations to determine the
main-sequence masses by averaging, for each star, the mass determined from
interpolation of the stellar temperature and that from interpolation of
the stellar luminosity.  Uncertainties are estimated from the uncertainties 
in the stellar properties (luminosity and temperature) and the
difference between the luminosity-predicted and temperature-predicted
masses.  This procedure could not be followed for the PS99 tracks since no
10$^9$ year isochrone exists and the 10$^8$ year isochrone exists only in
the mass range 0.1-0.8 M$_\odot$; no main-sequence masses are determined
from these models.

The validity of our adopted main-sequence isochrone merits some discussion.
Since there is continuous luminosity and temperature evolution even when
stars are on the main sequence, our derived masses are appropriate, in a
strict sense, only for the specific age assumed in creating the
mass-luminosity or mass-temperature relationships.  For example, at masses
above 0.7 M$_\odot$ where we have adopted the relationships for 10$^8$
years, a 1.0 M$_\odot$ star will have its mass overestimated by 2\% if it is
really 10$^9$ years old while a 2.0 M$_\odot$ star will have its mass
overestimated by 10\%.  One might think about assuming for all stars in our
main sequence sample the mean age in the solar neighborhood of $\sim$3 Gyr.
This approach would be incorrect, however, since we have selected stars via
their surface gravity to be on the hydrogen-burning main sequence,
which corresponds to different mean ages at different masses.  If a star is
really 3$\times10^9$ years old it will not be in our main sequence sample
at 2.0 M$_\odot$, but at 1.0 M$_\odot$ will have its mass overestimated by
6\%.  Without precise knowledge of the ages of the stars in our sample we
can only bear these biases in mind; we can not correct for them.
Because the hydrogen-burning main sequence is widest for the most massive
($>$10 M$_\odot$) stars and decreases in width towards lower masses, this
effect should not limit the conclusions drawn from our primarily low mass
sample.

\section{Comparison of Track Predicted Masses to Dynamical
Masses}\label{sec_comp}  

\begin{figure}
\epsscale{0.55}
\plotone{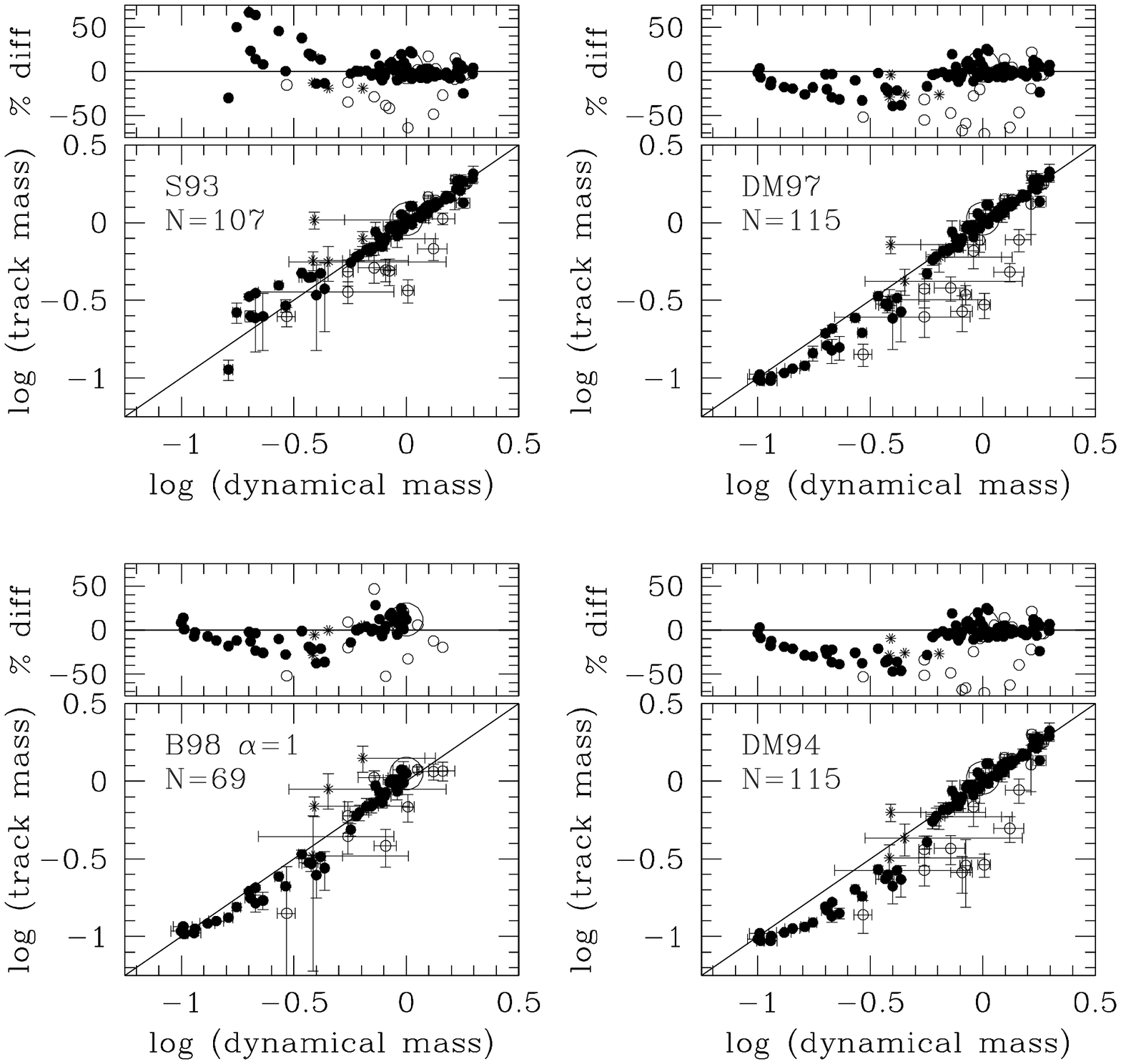}
\plotone{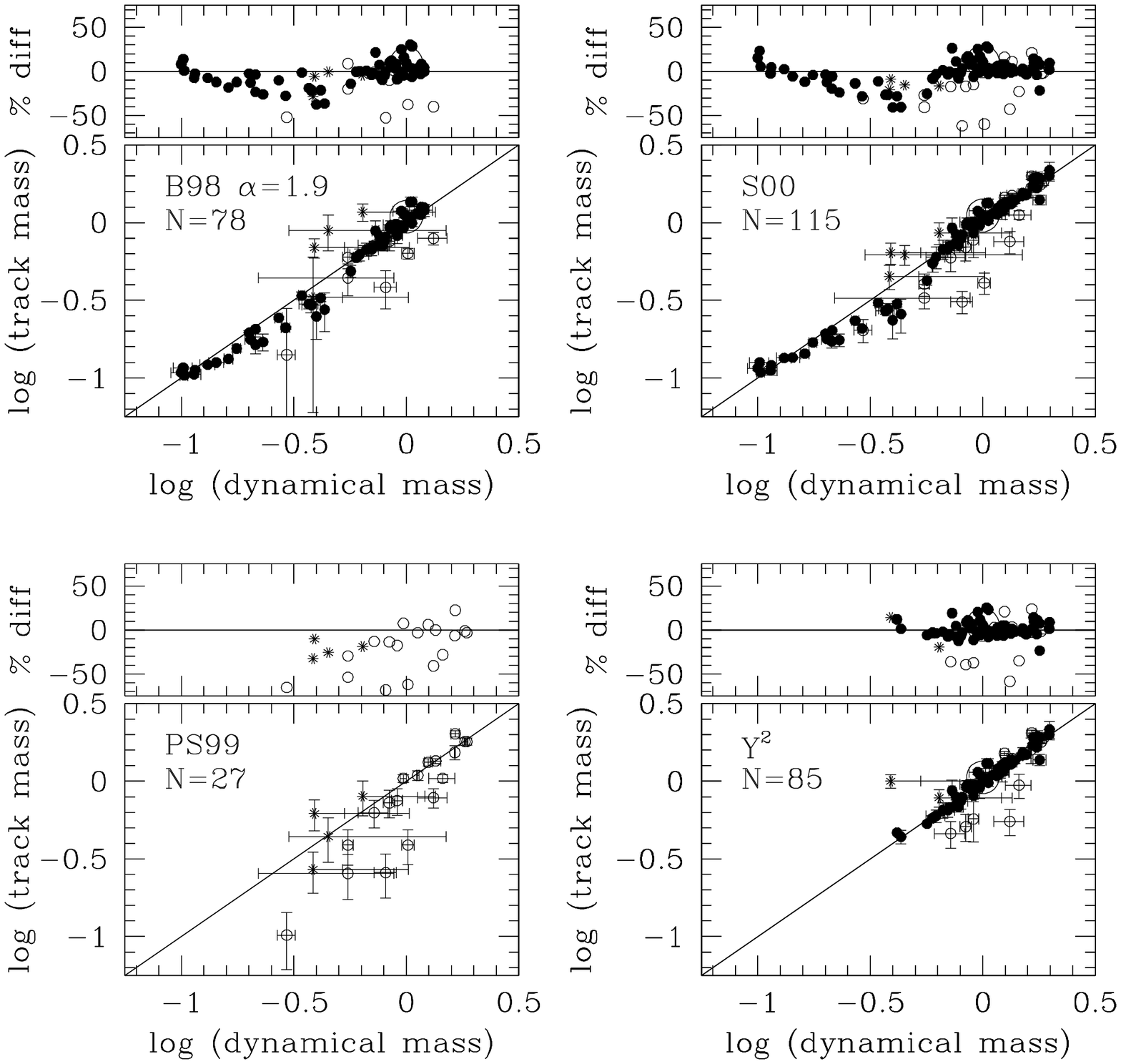}
\epsscale{1.0}
  \caption{
  Comparison of dynamically determined masses with track-predicted masses
  in units of solar masses
  for main sequence (filled symbols) and pre-main sequence (open symbols)
  stars. Asterisks represent pre-main sequence binary systems whose individual 
  components can be placed in the HR diagram but whose 
  measured dynamical mass is that of the composite system; these systems
  have been plotted assuming an average dynamical mass of 1/2 the total
  dynamical mass and an average percentage mass difference of 1/2 the total. 
  The Sun is also shown as the large open circle.  
  Not all stars in Table 1 appear in all panels due to the variation between
  model calculations in the range of masses covered.  The percentage
  mass difference in the upper panels is in the sense of track-predicted 
  mass minus dynamical mass.  
  \label{comparem} }
\end{figure}

\begin{figure}
\epsscale{1.0}
\plotone{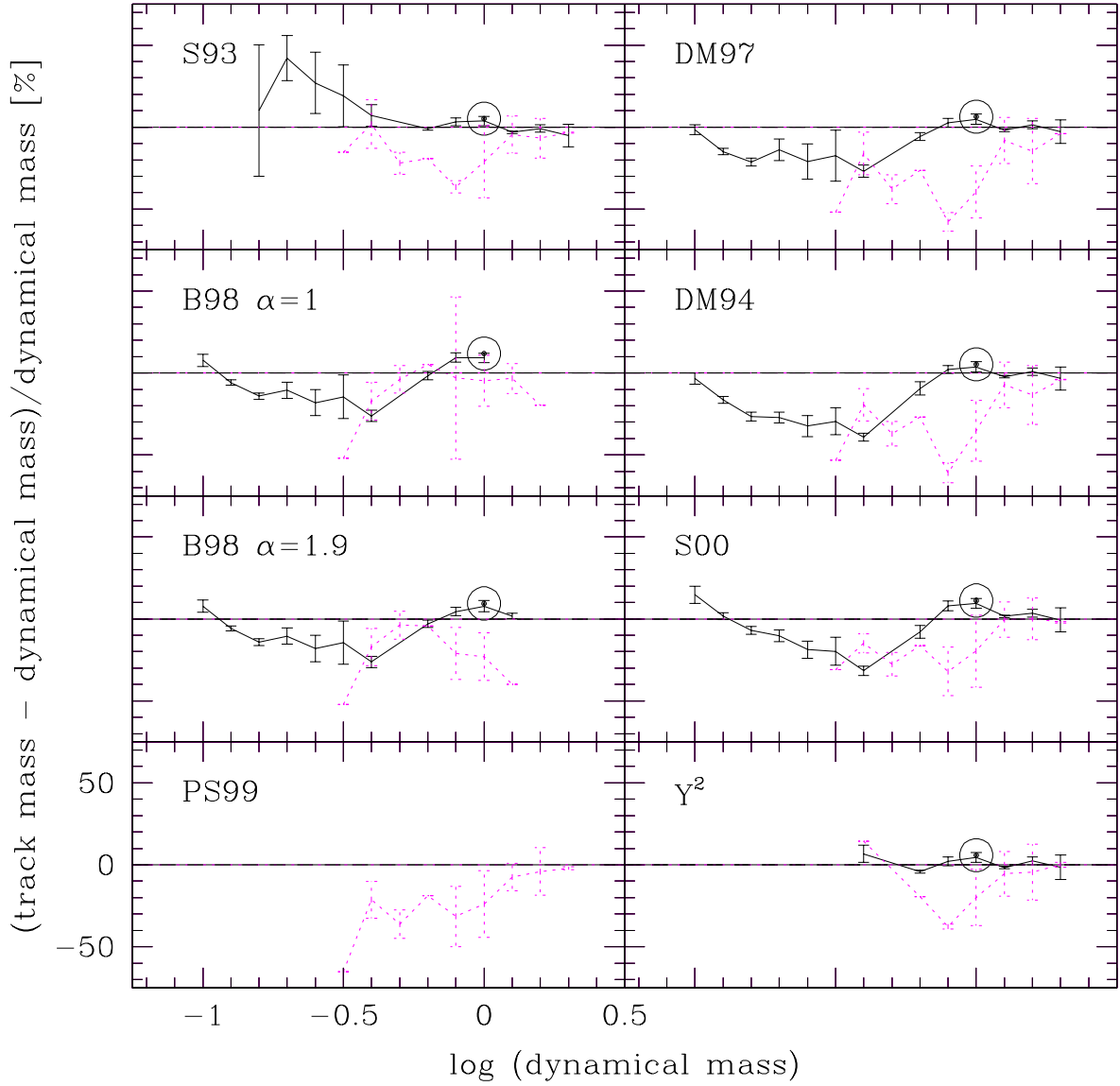}
  \caption{
  Mean percentage mass offset as a function of dynamically determined
  stellar mass for main sequence (solid lines)
  and pre-main sequence (dotted lines) stars; vertical error bars
  indicate the standard deviation of the mean.
  The difference values for the 4.5 Gyr old Sun are
  also shown as the large solar symbol.  
  \label{comparemav} }
\end{figure}

Figure \ref{comparem} shows comparisons between the dynamically determined
masses and
the masses inferred from all eight sets of evolutionary tracks; both the
direct correlation of mass and the difference between the two masses as a
percentage of the dynamical mass are provided.  Figure \ref{comparemav}
shows the mean percentage differences between track-predicted and dynamical
masses as a function of dynamically determined mass (essentially a binned
version of the upper plots in Figure \ref{comparem}).  The standard
deviations of the means are plotted as error bars for statistical
assessment.  In both of these Figures 
the main-sequence and the pre-main sequence samples are distinguished.  
The binary systems which have only total system
dynamical masses (FO Tau A/B, FS Tau A/B, DF Tau A/B, GG Tau Aa/b) have
been plotted assuming that the average mass per star is 1/2 the total
dynamical mass and that the average offset per star is 1/2 the total system
difference.  This assumption is justified given the similar spectral types
of the components of these binaries (Table 1).  Figures \ref{comparem} and
\ref{comparemav} illustrate the differences between the
predictions of the various pre-main sequence evolutionary calculations and
are now used to assess the robustness of the predicted stellar masses.  

\subsection{Main Sequence Stars}

We first consider the comparison of the main sequence sample.  For the 5
tracks that extend to the largest masses considered here, 1.2 - 2.0
M$_\odot$ (S93, DM94, DM97, S00, Y$^2$), there is excellent agreement
between the theoretical and dynamically determined masses in all cases.
Closer to 1.0 M$_\odot$, the S93, DM94, DM97 and Y$^2$ models again
predict main sequence masses that are consistent with dynamically
determined values.  However, both B98 models and the S00 models predict
masses that are 5\% (at 1-2 $\sigma$) larger than the dynamical masses.
This could be an evolutionary effect since the average age of the solar
mass main-sequence stars in our sample is likely more than 10$^8$
years.  Note, however, that the Sun (indicated by the solar symbol) resides
just beyond the one-sigma error in the mean difference, likely indicating
that the Sun is slightly older than the mean 1 M$_\odot$ star in our
sample; as discussed in \S 3.4, the $10^8$yr isochrone will overestimate
the mass of the Sun by 6\%, roughly the magnitude of the observed offset.  
At sub-solar masses, all tracks except for S93 and Y$^2$ predict masses that 
are less than the dynamically determined values by 15-30\% at several
sigma significance.  The Y$^2$ models show the flattest overall trend
(see Figure \ref{comparemav}) with agreement between predicted and
dynamically determined masses to within 1-3\% over all masses down to 0.6
M$_\odot$; the agreement slips to 7\% for the lowest considered
mass of 0.4 M$_\odot$.  The S93 models, in contrast 
to other models at low masses, are consistent with dynamical masses down to
0.3 M$_\odot$ but systematically  {\it over-predict} (as opposed to
under-predict) the lower masses.  Near 0.1 M$_\odot$ (the 2 lowest mass
bins), all models that extend this low appear to reverse their
offset trends and again predict masses that are consistent with the
dynamically inferred values.  

The systematic discrepancy of predicted and dynamical masses for  
0.2 - 0.5 M$_\odot$ main-sequence stars likely stems from the poor
match of model 10$^9$ year isochrones (our adopted main-sequence over
this mass range) with the empirical main-sequence, as shown in Figure
\ref{zamshrd}.  We note that this empirical main sequence is consistent with 
the location of low mass main-sequence members of our sample (Figure 3),
confirming that these stars are not peculiar because of, for example,
chromospheric activity.  The DM97, B98, and S00 models, which all
under-predict low-mass stellar masses, are either too hot by $\sim 200$
degrees or under-luminous by a factor of three.  We note though that 
masses determined via interpolation of stellar luminosity are more
consistent with dynamically determined values than the masses determined
via interpolation of stellar temperature (the adopted values for comparison
to dynamical masses are the average of the luminosity-predicted and 
the temperature-predicted masses; see \S 3.4).  This suggests that the 
main source of discrepancy in the models is in temperature and not the 
luminosity. 

A major cause of systematic disagreement between low-mass dynamical masses
and track-predicted masses is disparity between observation and theory in
the ``break" in the mass-luminosity
relationship (seen in the figures as a break in the temperature-luminosity
relationship).  In most models this break occurs at a hotter 
temperature (log T $\sim$ 3.7 dex; M0.5 spectral type) than the location of the
empirical break (log T $\sim$ 3.5 dex; M3.5 spectral type).  Even the Y$^2$
models, which predict the most consistent masses, are clearly diverging
from the empirical main-sequence over this mass range; these models exhibit
no break in their mass-luminosity (temperature-luminosity) relationship.
Only the S93 models offer reasonable agreement with the empirical
main-sequence at low masses. Interestingly, the standard deviation of the
mean offset is much larger at low masses for the S93 models than for other
models; this is because the data scatter uniformly around this main
sequence whereas for other models the offset between the data and the
predicted main sequence is large, and the standard deviation in the mean
offset is substantially smaller since {\it all} the data are offset in the
same direction and by roughly the same amount.  Similar conclusions
regarding the accuracy of the predicted main sequence can be derived by
comparing open cluster loci to these models (e.g. Stauffer et al. 1995;
Hillenbrand et al. 2004).  At high masses, the divergence seen in
Figure~\ref{zamshrd} between the models and the empirical main sequence is
expected since most main sequence stars (i.e. those used to derive the
absolute and bolometric magnitudes of typical main sequence stars) are
slightly more evolved than the theoretical ZAMS.

\subsection{Pre-Main Sequence Stars}

\begin{figure}
\epsscale{1.0}
\plotone{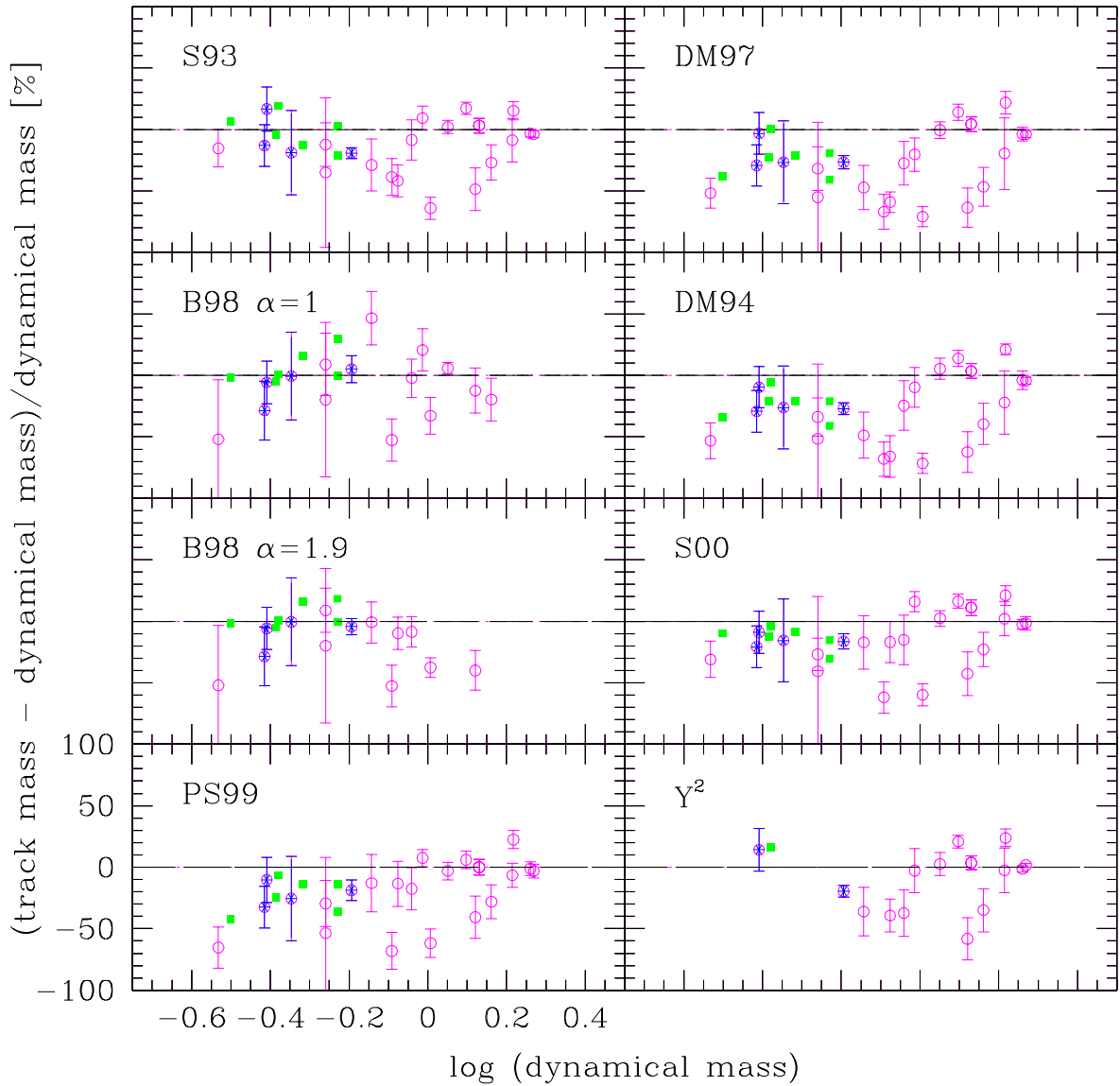}
  \caption{
  Percentage mass offset versus dynamically determined
  stellar mass for individual pre-main sequence stars. Vertical error bars
  indicate the root-sum-squared of the dynamical mass error and the track
  mass error, the latter estimated from the log L and log T errors.
  To illustrate the effects of temperature scale choice we show both
  the dwarf temperature scale adopted here (open circles) and the warmer 
  Luhman (2003) temperature scale (filled squares) for stars later than M0,  
  offset by $+$0.03 in log dynamical mass for clarity.
  Note the change in scale compared to Figure~\ref{comparemav}.
  \label{comparempms} }
\end{figure}

We now consider the pre-main sequence sample.   Relative to our main sequence 
sample these stars have poorly constrained temperatures and luminosities, 
leading to larger errors in HR diagram placement and hence larger errors 
in predicted masses.
In addition, the errors in the dynamical masses for this sample are
often substantially larger than the 10\% limit we imposed on the main sequence
sample.  Finally, the statistics
for the pre-main sequence are comparatively worse given the small number
of pre-main sequence stars with dynamically determined masses.  With these
caveats in mind, we interpret the comparisons shown in Figures 
\ref{comparem} and \ref{comparemav} with the aid 
of Figure~\ref{comparempms} which shows the results for individual stars,
similar to the top panels of Figure~\ref{comparem}, but with an expanded scale
and now with individual error bars.

Above 1.2 M$_\odot$, all models considered (all except both B98 calculations
which do not extend above this mass) predict pre-main sequence 
masses that are consistent with dynamically determined values 
to better than 1$\sigma$ in the mean (Figure~\ref{comparemav}),
with the DM94 and DM97 tracks tending to
under-predict the individual masses by 0-10\%.  
Around 1 M$_\odot$ (0.5-1.2 M$_\odot$), 
the B98 $\alpha$=1.0 models predict masses most consistent with dynamical
values; the B98 $\alpha$=1.9 and most other models predict masses that 
are too low by $\sim 25$\% at $1-2\sigma$ on average compared to  
the dynamically determined values.  This general trend of underpredicted
masses continues (including for the B98 $\alpha$=1.0 models)
towards the lowest pre-main sequence masses considered, 0.3 M$_\odot$ though 
with slightly less significance ($\sim 1 \sigma$).  Note that the
valley of maximum disagreement between track-predicted and dynamical masses
is driven for all models by two stars: UZ Tau Aa and NTTS 045251 B.   

Our assessment of these mass comparisons is limited by the accuracy 
with which our sample stars can be placed on an H-R diagram, particularly
the youngest stars.  As young solar- and lower mass stars are primarily on
Hayashi (roughly constant temperature) evolutionary tracks, an accurate
temperature is especially important for determining a theoretical mass.  In
our analysis we have adopted a dwarf temperature scale for both the main
sequence stars and the pre-main sequence stars.  Pre-main sequence stars are
intermediate gravity objects between dwarfs and giants and it has been 
argued (e.g. Mart\'\i n et al. 1994; Luhman et al. 1997; White et al. 1999)
that the appropriate spectral type - temperature relation of, in particular
T Tauri stars, should be intermediate between that of dwarfs and giants. 
G and K giants are cooler than G and K dwarfs, while M giants are warmer 
than M dwarfs (see Appendix for dwarf temperatures and Dyck et al. 1996, 
Di Benedetto \& Rabbia 1987, and Bell \& Gustaffson 1989 for giant temperatures
derived from either angular diameters or the infrared flux method) 
with the crossover point at about M0.  As examples, in comparison to dwarfs,
giants of spectral type M6 are $\sim$620 K warmer, M4 are $\sim$500 K warmer, 
M2 are $\sim$310 K warmer, K5 are $\sim$475 K cooler, and K1 are $\sim$595 K cooler. 
Detailed analysis of high dispersion spectra shows that pre-main sequence
surface gravities are closer to dwarfs than giants.  For example,
Johns-Krull et al. (1999) measure log g = 3.67 $\pm$0.5 for BP Tau and
Johns-Krull \& Valenti (2001) quote log g = 3.54 for Hubble 4.  These
values can be compared to log g = 4.6 for a 4800K dwarf and log g = 2.4 for
a 4800 K giant (dwarf surface gravities staying roughly constant with
decreasing temperature in the stellar range and giant gravities decreasing by
one order of magnitude by 3900 K and two orders of magnitude by late M
spectral types).  

In our analysis, we have assumed a strict dwarf-like
temperature relation since an appropriate temperature scale tied to
the infrared flux method or measured stellar angular diameters has not yet
been established for 1-10 Myr low mass stars.  The systematic shift induced
by adopting a temperature scale intermediate to that of dwarfs and giants
would make our track-inferred masses for the pre-main sequence stars 
{\it smaller} in
the GK spectral type range (the wrong direction for improving correspondance
to dynamical masses) and {\it larger} by $\sim$10\% for the
M types.  Luhman et al. (2003) suggest a specific intermediate temperature
scale for stars cooler than spectral type M0\footnote{The values of the
Luhman intermediate temperature scale were chosen to produce co-eval ages
for the T Tauri quadruple GG Tauri and for members of the IC 348 cluster
using the B98 ($\alpha$ = 1.9) evolutionary models.}.
Using this warmer temperature scale for our pre-main sequence sample 
(filled squares in Figure~\ref{comparempms})  
systematically increases the predicted masses of the lowest mass stars.
However there is no statistically significant evidence
from dynamical mass constraints that a warmer-than-dwarf temperature scale
is needed since the resulting change in the predicted masses using a warmer
scale is well within the uncertainties in the mass comparison plots
(only 2 systems have masses shifted by $\ge 1\sigma$
via a change in the temperature scale). 

Systematic shifts in the predicted masses, as would occur by shifting the
temperature scale, will still leave many pre-main sequence stars with 
track-predicted masses widely discrepant from dynamical values.  This is
illustrated by the large scatter in track predicted masses over a small
range of dynamically determined masses (Figure ~\ref{comparempms}).  
A couple of case studies make this point clear.  Compare MWC 480, an
A2 star with dynamical mass of $1.65\pm0.07$ M$_\odot$, to the cooler but
(surprisingly) more massive A8 stars RS Cha A \& B, with dynamical  
masses of $1.858\pm0.016$ M$_\odot$ and $1.821\pm0.018$ M$_\odot$.  No 
evolutionary model will predict that a hotter object is less massive than a
cooler object this close to the main-sequence.  Assuming that the
uncertainties in the dynamical masses have been properly assigned, this
suggests that the assigned temperatures are in error.  In this case, the
error is most likely in the spectral type assigned to MWC 480 since RS Cha
is an eclipsing system with more precisely determined temperatures.
Similar discrepancies occur at lower masses.  Consider NTTS 04251+3016 A
and LkCa 15, two K5 T Tauri stars with identical luminosities.  Although
these stars are located at the same position in the H-R diagram, they have
dynamically determined masses that 
differ by 0.48 M$_\odot$, a 2.5$\sigma$ difference.  This again strongly
suggests errors in the assigned spectral types.  These discrepancies are
problems that will remain, independent of the temperature scale and
independent of any evolutionary model.  Assuming that the uncertainties in
dynamical masses are being properly assessed, we conclude that
  usefulness of dynamical mass constraints on pre-main sequence 
evolutionary models is currently limited by poorly determined
luminosities and especially temperatures of pre-main sequence stars.

\subsection{Ensemble Comparisons}

Finally, in assessing the main sequence and the pre-main sequence results
en ensemble, we find it somewhat distressing that for most models the
agreement for main sequence masses is far better than for pre-main sequence
masses.  Assuming the stellar parameters on average
are well understood (the above exceptions notwithstanding), apparently it is 
possible for stars of given mass to wind up in the right place near the main
sequence end of a calculation without having started them in the right place 
at the tops of their convective evolutionary tracks.  

The B98 ($\alpha$=1.0)
models appear to have the best consistency between the pre-main sequence
and main sequence mass offsets as a function of mass (Figure ~\ref{comparemav}), 
though we remind the
reader that we found the B98 $\alpha$=1.0 models a better fit to the
pre-main sequence and the B98 $\alpha$=1.9 models a better fit to the
main sequence.  If this trend is proven true, it may indicate a difference
in the efficiency of convection between pre-main sequence and main sequence
stars of similar mass.  As noted above, for all models there is indeed
{\it consistency} above 1.2 M$_\odot$ in both the pre-main sequence and the main
sequence phases with dynamical masses; however, the pre-main sequence
masses are systematically offset by 0-30\% ($<1\sigma$). Below 1 M$_\odot$ 
the consistency between the pre-main sequence and main sequence
masses is broken, with the offset masses in the two regimes different in most 
models by $>1\sigma$.  Notably it is in this sub-solar regime where  
convection is most important, and for an increasingly longer time period 
towards lower masses, during pre-main sequence evolution.  

\section{Conclusions and Recommendations}\label{sec_conc}

We have attempted to assess the agreement between
dynamically determined stellar masses and those inferred from modern theoretical
calculations of pre- and early-main sequence evolution.  We have found
only marginal consistency with most existing models, as summarized in 
Figure \ref{comparemav}. 

For main-sequence stars, above 1.2 M$_\odot$
the models considered are all consistent with dynamically determined values.  
At lower masses, however, there is divergence between the predicted and
dynamical masses which sets in at different masses for different tracks.
The Y$^2$ models offer the best overall agreement with dynamical
masses, though these calculations extend only as low in mass as 0.4 M$_\odot$.  
The S93 models are a close second to the Y$^2$ models but begin to diverge 
from 1$\sigma$ consistency below 0.3 M$_\odot$.  All other models
(DM97, B98, S00, and PS99)
fail to predict masses that are consistent with dynamically determined
values (by 5-20\%) over the mass range 0.1 - 0.5 M$_\odot$.  
We find that for all tracks, the
dominant discrepancies between track-predicted and dynamically determined
masses for main-sequence stars lie in the mass range 0.2 to 0.5
M$_\odot$.  This failure
likely stems from the poor match to the empirically defined main-sequence.
The DM97, B98, PS99, and S00 models all predict a break in the mass-luminosity
relationship near log T $\sim$ 3.7 dex (spectral type M0.5), which is
hotter than the well established empirical break in the mass-luminosity
relationship near log T $\sim$ 3.5 dex (spectral type M3.5).
The S93 and Y$^2$ models most closely resemble the empirical main sequence.

For the pre-main sequence sample, we find generally good agreement between 
predicted and dynamical masses above 1.2 M$_\odot$ for all models, as was
true for the main sequence sample.  This is not an entirely trivial statement
since both partially convective and fully radiative stars are included
between these two samples.  However, referring to Figure~\ref{comparetrack},
differences between the various models for 1-2 M$_\odot$ stars 
are manifest only high on the fully convective part of the tracks where no
empirical data exists; thus even younger 1-2 M$_\odot$ dynamical masses 
are needed before distinction between the pre-main sequence 
tracks can be made in this mass regime.
Between 1.2 and 0.5 M$_\odot$, the B98 ($\alpha$=1.0) 
models predict {\it reasonably} though 
not fully consistent mass values on average, while all other models  
systematically underestimate sub-solar masses by 10-30\% on average.  
At the lowest masses considered, $\lesssim 0.5$ M$_\odot$, all
models underestimate pre-main sequence stellar masses. There are at present no
dynamical mass constraints available at masses less than 0.3 M$_\odot$ 
for pre-main sequence stars.
Adopting a warmer-than-dwarf temperature scale for T Tauri
stars could partly reconcile these mass under-estimates, though the
scale proposed by Luhman et al. (2003) is not warm enough to rectify
the mass underestimates except for the marginal (that is, not statistically
significant) improvements made to the B98 model agreement (the models to
which this temperature scale was in fact tuned).  With the above
caveats in mind, we find that the B98 $\alpha$=1.0 models used with a slightly 
warmer-than-dwarf temperature scale predict pre-main sequence masses 
that agree the most consistently with dynamically determined values.  
Of note is that the B98 models do not extend above radii of 1-2 R$\odot$
(specificially the 10$^6$ year isochrone) whereas many young pre-main sequence 
stars have larger radii, 2-3 R$\odot$, thus limiting the utility of the B98 
models in star-forming regions.  The dynamical mass consistency of the 
B98 models is only marginally better than the DM97, PS99, and S00 models, 
which systematically underestimate sub-solar masses by 1-2$\sigma$.

The relatively flat nature
of the offsets between the dynamical and the predicted stellar masses for
some calculations suggests that they could be used with moderate confidence
if correction factors are included.  For example, a 20\% revision upward
of the masses predicted by the DM97 tracks for masses between 
0.12-0.4 M$_\odot$ would result in near-perfect agreement at 
main sequence evolutionary stages, with the same 20\% correction 
applicable to 0.3-1.0 M$_\odot$ young pre-main sequence stars; again we note 
that the pre-main sequence behavior below 0.3 M$_\odot$ is untested for these 
or any set of tracks.  
A similar 20\% correction could be applied to the S00 pre-main sequence tracks,
though the main sequence offsets appear to vary with mass.


Several observational recommendations can also be made.  Our pre-main sequence 
comparisons stress the need for more observational work on masses determined
from orbital dynamics in the 
pre-main sequence phase where the statistics of our assembled sample 
are factors of 5-10 worse than on the main sequence at comparable masses.
This is especially problematic at the lowest masses where at present
there are no pre-main sequence dynamical mass constraints 
at masses $<$0.3 M$_\odot$.  Finally, we emphasize that the usefulness of
dynamical mass constraints on pre-main sequence evolutionary 
models are currently limited
by poorly determined luminosities and especially temperatures of young
stars.  Additional dynamical mass determinations will not likely improve the
constraints on evolutionary models unless the stellar parameters can be more
accurately determined than for the current sample.  
In the absence of additional eclipsing systems,
high dispersion stellar spectroscopy and synthetically modeled spectra
offer the best promise for precisely determining fundamental properties.


The trends that have emerged from our study may be interpretable as
messages regarding modifications to the model assumptions on input physics and
parameter choices.  
It is suggested that in order to achieve agreement between dynamical and 
track-predicted masses for both low-mass young pre-main sequence stars and
main sequence stars, a systematic shift coolward of the models via improved 
convection and opacity treatments are needed.  Further adjustments may also be 
necessary.  Baraffe et al. have repeatedly stressed the important
effects of atmospheres at low masses, arguing that the grey (Eddington)
approximation used by most other authors overestimates both the temperature
and the luminosity for a given mass.  This in part could explain some of
the discrepancies between the predicted and empirical main-sequences
(Figure \ref{zamshrd}).  It is worth noting that the deviations occur near
early M spectral types, where molecular absorption begins to dominate the
opacity.  However, even the non-grey atmospheres of the B98 models fail to
reproduce the empirical main-sequence.  For the pre-main sequence stars,  
although the physics involved in opacities, equations of state, and
atmospheric treatment is already challenging, even more sophisticated
effects such as accretion, rotation, and magnetic fields may be required
in order to achieve rigorous agreement between observations and models,
as illustrated by e.g. D'Antona et al. 2000 and Baraffe et al. 2002.


\acknowledgments
We acknowledge useful comments by the referee.

\appendix
\section{Adopted Dwarf Temperatures and Bolometric Corrections}

As discussed in the text (\S 3.3 and 4.2), we have adopted 
a dwarf temperature scale based on the stellar temperatures of
Chlebowski \& Garmany 1991 (O3-O9); Humphreys \& McElroy 1984 (B0-B3);
Cohen \& Kuhi 1979 (B5-K6); Bessell 1991 (K7-M1); Wilking, Greene, Meyer 1999
(M2-M7.5); and Reid / Burgasser (M8-L-T).
Our bolometric corrections are those of:
Massey et al. 1989 (O3-B1); Code et al. 1976 (B2-G0); Bessell 1991 
and Bessell \& Brett 1988 (G0-M5); and Tinney, Mould, \& Reid 1993 (M6-M9,
converted from quoted values of BC$_K$).
The V-band bolometric corrections turn over
at spectral types later than late-G, 
and grow rapidly as flux shifts from the V-band into redder band passes.
I-band is generally the best 
wavelength at which to apply a bolometric correction for stars
in the early-K through mid-M spectral type range, both because the value of
the bolometric correction is small and because it is roughly constant 
with spectral type.  For very late M-types the J-band may be a better choice.


\newpage
\pagestyle{empty}

\tablenum{1}
\begin{deluxetable}{l c c c l l l l c c l c l}
\rotate
\tabletypesize{\scriptsize}
\tablewidth{675pt}
\tablecaption{Sample and Stellar Parameters}
\tablehead{
\colhead{ Name }&
\colhead{ M/M$_\odot$ }&
\colhead{ R/R$_\odot$}&
\colhead{ log g/cm~s$^{-2}$}&
\colhead{ Type \tablenotemark{a}}&
\colhead{ ref.}&
\colhead{ SpT}&
\colhead{ ref.}&
\colhead{ log T$_{eff}$ \tablenotemark{b}}&
\colhead{ log L/L$_\odot$}&
\colhead{ ref.}&
\colhead{ Ev. \tablenotemark{c}}&
\colhead{ Comment}\\
}
\startdata
\hline
\multicolumn{12}{c}{Candidate Main Sequence Stars} \\
\hline
  WW Aur A &    1.987 $\pm$    0.034 &   1.883 $\pm$   0.038 &   4.187 $\pm$   0.019 &      EB &      A91 &      A5m&       A91 &    3.910$\pm$     0.015 &    1.140$\pm$     0.060 &      A91 &    3 &                                                   \\ 
V909 Cyg A &    1.980 $\pm$    0.030 &   1.470 $\pm$   0.020 &   4.403 $\pm$   0.012 &      EB &     L97c &       A0&      L97c &    3.987$\pm$     0.021 &    1.230$\pm$     0.090 &     L97c &    1 &                                                   \\ 
  KW Hya A &    1.978 $\pm$    0.036 &   2.125 $\pm$   0.016 &   4.079 $\pm$   0.013 &      EB &      A91 &      A5m&       A91 &    3.900$\pm$     0.006 &    1.220$\pm$     0.040 &      R03 &    3 &                                                   \\ 
  AI Hya B &    1.978 $\pm$    0.036 &   2.766 $\pm$   0.017 &   3.850 $\pm$   0.010 &      EB &      A91 &      F0V&       A91 &    3.869$\pm$     0.010 &    1.312$\pm$     0.036 &      R03 &    3 &                                                   \\ 
V1647 Sgr B &   1.972 $\pm$    0.033 &   1.666 $\pm$   0.017 &   4.289 $\pm$   0.012 &      EB &      A91 &      A1V&       A91 &    3.949$\pm$     0.014 &    1.192$\pm$     0.057 &      R03 &    1 &                                                   \\ 
  TZ For B &    1.949 $\pm$    0.027 &   3.962 $\pm$   0.088 &   3.532 $\pm$   0.020 &      EB &      A91 &     F7IV&       A91 &    3.803$\pm$     0.007 &    1.360$\pm$     0.030 &      A91 &    3 &                                                   \\ 
V624 Her B &    1.881 $\pm$    0.013 &   2.209 $\pm$   0.034 &   4.024 $\pm$   0.014 &      EB &      A91 &      A7V&       A91 &    3.900$\pm$     0.008 &    1.240$\pm$     0.040 &      A91 &    3 &                                                   \\ 
  MY Cyg B &    1.811 $\pm$    0.025 &   2.193 $\pm$   0.050 &   4.014 $\pm$   0.021 &      EB &      A91 &      F0m&       A91 &    3.846$\pm$     0.010 &    1.019$\pm$     0.045 &      R03 &    3 &                                                   \\ 
  GK Dra B &    1.810 $\pm$    0.109 &   2.830 $\pm$   0.054 &   3.790 $\pm$   0.041 &      EB &      Z03 &  \nodata&    \nodata&    3.837$\pm$     0.004 &    1.188$\pm$     0.029 &      Z03 &    3 &                                                   \\ 
  51 Tau A &    1.800 $\pm$    0.130 &   \nodata             &   \nodata             &       O &      T97a &  \nodata&   \nodata &    3.859$\pm$     0.013 &    1.046$\pm$     0.040 &      T97a &    3 &                                             Hyades member \\ 
  WW Aur B &    1.799 $\pm$    0.025 &   1.883 $\pm$   0.038 &   4.143 $\pm$   0.018 &      EB &      A91 &      A7m&       A91 &    3.890$\pm$     0.015 &    1.060$\pm$     0.060 &      A91 &    3 &                                                   \\ 
V477 Cyg A &    1.790 $\pm$    0.120 &   1.570 $\pm$   0.050 &   4.300 $\pm$   0.030 &      EB &     GQ92 &  \nodata&    \nodata&    3.939$\pm$     0.015 &    1.100$\pm$     0.050 &     MQ92 &    1 &                                                   \\
V477 Cyg B &    1.790 $\pm$    0.120 &   1.270 $\pm$   0.040 &   4.360 $\pm$   0.030 &      EB &     GQ92 &  \nodata&    \nodata&    3.826$\pm$     0.015 &    0.470$\pm$     0.050 &     MQ92 &    1 &                                                   \\ 
  MY Cyg A &    1.786 $\pm$    0.030 &   2.193 $\pm$   0.050 &   4.008 $\pm$   0.021 &      EB &      A91 &      F0m&       A91 &    3.850$\pm$     0.010 &    1.035$\pm$     0.045 &      R03 &    3 &                                                   \\ 
V909 Cyg B &    1.750 $\pm$    0.030 &   1.570 $\pm$   0.030 &   4.288 $\pm$   0.017 &      EB &     L97c &       A2&      L97c &    3.944$\pm$     0.016 &    1.120$\pm$     0.070 &     L97c &    1 &                                                   \\ 
  IQ Per B &    1.737 $\pm$    0.031 &   1.503 $\pm$   0.017 &   4.323 $\pm$   0.013 &      EB &      A91 &      A6V&       A91 &    3.906$\pm$     0.008 &    0.850$\pm$     0.040 &      R03 &    1 &                                                   \\ 
  OO Peg A &    1.720 $\pm$    0.030 &   2.190 $\pm$   0.080 &   3.990 $\pm$   0.040 &      EB &      M01 &  \nodata&    \nodata&    3.943$\pm$     0.007 &    1.388$\pm$     0.044 &      M01 &    3 &                            \\ 
  OO Peg B &    1.690 $\pm$    0.030 &   1.370 $\pm$   0.050 &   4.390 $\pm$   0.040 &      EB &      M01 &  \nodata&    \nodata&    3.939$\pm$     0.009 &    0.964$\pm$     0.048 &      M01 &    1 &                            \\ 
 V526 Sgr B&    1.680 $\pm$    0.060 &   1.560 $\pm$   0.020 &   4.280 $\pm$   0.020 &      EB &     L97b &  \nodata&    \nodata&    3.940$\pm$     0.005 &    1.100$\pm$     0.030 &     L97b &    1 &                                                   \\ 
 TV Nor B  &	1.665 $\pm$    0.018 &	 1.550 $\pm$   0.014 &	 4.278 $\pm$   0.012 &	    EB &     N97  &  \nodata&	 \nodata&    3.892$\pm$	    0.006 &    0.902$\pm$     0.035 &	  N97  &    1 &							  \\
  PV Pup A &    1.565 $\pm$    0.011 &   1.542 $\pm$   0.018 &   4.257 $\pm$   0.010 &      EB &      A91 &      A8V&       A91 &    3.840$\pm$     0.010 &    0.689$\pm$     0.041 &      R03 &    1 &                                                   \\ 
V442 Cyg A &    1.564 $\pm$    0.024 &   2.072 $\pm$   0.034 &   3.999 $\pm$   0.016 &      EB &      A91 &      F1V&       A91 &    3.839$\pm$     0.006 &    0.940$\pm$     0.030 &      R03 &    3 &                                                   \\ 
  PV Pup B &    1.554 $\pm$    0.013 &   1.499 $\pm$   0.018 &   4.278 $\pm$   0.011 &      EB &      A91 &      A8V&       A91 &    3.841$\pm$     0.010 &    0.668$\pm$     0.041 &      R03 &    1 &                                                   \\ 
  RZ Cha A &    1.518 $\pm$    0.021 &   2.264 $\pm$   0.017 &   3.909 $\pm$   0.009 &      EB &      A91 &      F5V&       A91 &    3.816$\pm$     0.010 &    0.926$\pm$     0.041 &      R03 &    3 &                                                   \\ 
  RZ Cha B &    1.509 $\pm$    0.027 &   2.264 $\pm$   0.017 &   3.907 $\pm$   0.010 &      EB &      A91 &      F5V&       A91 &    3.816$\pm$     0.010 &    0.926$\pm$     0.041 &      R03 &    3 &                                                   \\ 
  TZ Men B &    1.504 $\pm$    0.010 &   1.432 $\pm$   0.015 &   4.303 $\pm$   0.009 &      EB &      A91 &      A8V&       A91 &    3.857$\pm$     0.012 &    0.692$\pm$     0.070 &      R03 &    1 &                                                   \\ 
  KW Hya B &    1.488 $\pm$    0.017 &   1.480 $\pm$   0.014 &   4.270 $\pm$   0.010 &      EB &      A91 &      F0V&       A91 &    3.836$\pm$     0.007 &    0.637$\pm$     0.029 &      R03 &    1 &                                                   \\ 
  BW Aqr A &    1.488 $\pm$    0.022 &   2.064 $\pm$   0.044 &   3.981 $\pm$   0.020 &      EB &      A91 &      F7V&       A91 &    3.800$\pm$     0.007 &    0.782$\pm$     0.034 &      R03 &    3 &                                                   \\ 
  GK Dra A &    1.460 $\pm$    0.066 &   2.431 $\pm$   0.042 &   3.830 $\pm$   0.033 &      EB &      Z03 &  \nodata&    \nodata&    3.851$\pm$     0.004 &    1.112$\pm$     0.030 &      Z03 &    3 &                                                   \\ 
  DM Vir A &    1.454 $\pm$    0.008 &   1.763 $\pm$   0.017 &   4.108 $\pm$   0.009 &      EB &      L96 &      F7V&       A91 &    3.806$\pm$     0.010 &    0.700$\pm$     0.030 &      R03 &    3 &                                                   \\ 
  DM Vir B &    1.448 $\pm$    0.008 &   1.763 $\pm$   0.017 &   4.106 $\pm$   0.009 &      EB &      L96 &      F7V&       A91 &    3.806$\pm$     0.010 &    0.700$\pm$     0.030 &      R03 &    3 &                                                   \\ 
  CD Tau A &    1.442 $\pm$    0.016 &   1.798 $\pm$   0.017 &   4.087 $\pm$   0.010 &      EB &      R99 &      F6V&       R99 &    3.792$\pm$     0.004 &    0.630$\pm$     0.020 &      R99 &    3 &                                                   \\ 
  AD Boo A &    1.438 $\pm$    0.016 &   1.614 $\pm$   0.012 &   4.180 $\pm$   0.011 &      EB &      L97 &  \nodata&    \nodata&    3.805$\pm$     0.006 &    0.590$\pm$     0.030 &      L97 &    3 &                                                   \\ 
V442 Cyg B &    1.410 $\pm$    0.023 &   1.662 $\pm$   0.033 &   4.146 $\pm$   0.019 &      EB &      A91 &      F2V&       A91 &    3.833$\pm$     0.006 &    0.720$\pm$     0.030 &      R03 &    3 &                                                   \\ 
V1143 Cyg A &   1.391 $\pm$    0.016 &   1.346 $\pm$   0.023 &   4.323 $\pm$   0.016 &      EB &      A91 &      F5V&       A91 &    3.820$\pm$     0.008 &    0.491$\pm$     0.035 &      R03 &    1 &                                                   \\ 
  BW Aqr B &    1.386 $\pm$    0.021 &   1.788 $\pm$   0.043 &   4.075 $\pm$   0.022 &      EB &      A91 &      F8V&       A91 &    3.807$\pm$     0.007 &    0.685$\pm$     0.035 &      R03 &    3 &                                                   \\ 
  CD Tau B &    1.368 $\pm$    0.016 &   1.584 $\pm$   0.020 &   4.174 $\pm$   0.012 &      EB &      R99 &      F6V&       R99 &    3.792$\pm$     0.004 &    0.520$\pm$     0.020 &      R99 &    3 &                                                   \\ 
  YZ Cas B &    1.350 $\pm$    0.010 &   1.348 $\pm$   0.015 &   4.309 $\pm$   0.010 &      EB &      A91 &      F2V&       A91 &    3.821$\pm$     0.016 &    0.496$\pm$     0.065 &      R03 &    1 &                                                   \\ 
V1143 Cyg B &   1.347 $\pm$    0.013 &   1.323 $\pm$   0.023 &   4.324 $\pm$   0.016 &      EB &      A91 &      F5V&       A91 &    3.816$\pm$     0.008 &    0.460$\pm$     0.035 &      R03 &    1 &                                                   \\ 
  EE Peg B &    1.335 $\pm$    0.011 &   1.312 $\pm$   0.013 &   4.328 $\pm$   0.009 &      EB &      A91 &      F5V&       A91 &    3.802$\pm$     0.005 &    0.396$\pm$     0.022 &      R03 &    1 &                                                   \\ 
  IT Cas A &    1.330 $\pm$    0.009 &   1.593 $\pm$   0.015 &   4.158 $\pm$   0.009 &      EB &     L97d &      F5V&      L97d &    3.811$\pm$     0.007 &    0.601$\pm$     0.035 &     L97d &    3 &                                                   \\ 
  IT Cas B &    1.328 $\pm$    0.008 &   1.560 $\pm$   0.040 &   4.175 $\pm$   0.020 &      EB &     L97d &      F5V&      L97d &    3.811$\pm$     0.007 &    0.583$\pm$     0.047 &     L97d &    3 &                                                   \\ 
V505 Per A &    1.300 $\pm$    0.020 &   1.400 $\pm$   0.020 &   4.260 $\pm$   0.010 &      EB &      M01 &  \nodata&    \nodata&    3.808$\pm$     0.003 &    0.456$\pm$     0.016 &      M01 &    1 &                                                   \\ 
V505 Per B &    1.280 $\pm$    0.020 &   1.140 $\pm$   0.030 &   4.430 $\pm$   0.010 &      EB &      M01 &  \nodata&    \nodata&    3.807$\pm$     0.004 &    0.280$\pm$     0.020 &      M01 &    1 &                                                   \\ 
V570 Per A &    1.280 $\pm$    0.030 &   1.640 $\pm$   0.160 &   4.120 $\pm$   0.050 &      EB &      M01 &  \nodata&    \nodata&    3.810$\pm$     0.010 &    0.600$\pm$     0.072 &      M01 &    3 &                                                   \\ 
  HS Hya A &	1.2552 $\pm$   0.0078 &	 1.2747 $\pm$  0.0072 &	 4.3259 $\pm$  0.0056 &	    EB &      T97b & \nodata&	 \nodata&    3.8129$\pm$    0.0033 &   0.415$\pm$     0.014 &	   T97b &   1 &							  \\
  RT And A &    1.240 $\pm$    0.030 &   1.260 $\pm$   0.015 &   4.335 $\pm$   0.015 &      EB &      P94 &      F8V&       S93 &    3.785$\pm$     0.015 &    0.290$\pm$     0.060 &      P94 &    1 &                                              CABS \\ 
  UX Men A &    1.238 $\pm$    0.006 &   1.347 $\pm$   0.013 &   4.272 $\pm$   0.009 &      EB &      A91 &      F8V&       A91 &    3.785$\pm$     0.007 &    0.351$\pm$     0.029 &      R03 &    1 &                                                   \\ 
  AD Boo B &    1.237 $\pm$    0.013 &   1.211 $\pm$   0.018 &   4.364 $\pm$   0.019 &      EB &      L97 &  \nodata&    \nodata&    3.775$\pm$     0.007 &    0.220$\pm$     0.040 &      L97 &    1 &                                                   \\ 
  AI Phe B &    1.231 $\pm$    0.005 &   2.931 $\pm$   0.007 &   3.593 $\pm$   0.003 &      EB &      M92 &      F7V&       A91 &    3.712$\pm$     0.013 &    0.730$\pm$     0.050 &      M92 &    3 &                                                   \\ 
  FL Lyr A &    1.221 $\pm$    0.016 &   1.282 $\pm$   0.028 &   4.309 $\pm$   0.020 &      EB &      A91 &      F8V&       A91 &    3.789$\pm$     0.007 &    0.320$\pm$     0.030 &      A91 &    1 &                                                   \\ 
V570 Per B &    1.220 $\pm$    0.030 &   1.010 $\pm$   0.025 &   4.550 $\pm$   0.120 &      EB &      M01 &  \nodata&    \nodata&    3.793$\pm$     0.013 &    0.080$\pm$     0.176 &      M01 &    1 &                                                   \\ 
  HS Hya B &	1.2186 $\pm$   0.0070 & 1.2161 $\pm$  0.0071 &	 4.3539 $\pm$  0.0057 &	    EB &      T97b & \nodata&	 \nodata&    3.8062$\pm$    0.0034 &   0.347$\pm$     0.015 &	   T97b	&   1 &  \\
  UV Leo A &    1.210 $\pm$    0.097 &   0.973 $\pm$   0.024 &   4.540 $\pm$   0.053 &      EB &      Z03 &  \nodata&       Z03 &    3.787$\pm$     0.005 &    0.060$\pm$     0.038 &      Z03 &    1 &                                                   \\ 
 HR 6697 A &    1.200 $\pm$    0.110 &   \nodata             &   \nodata             &       O &      P00 &      G0V&      Mc95 &    3.771$\pm$     0.015 &    0.211$\pm$     0.040 &     Mc95 &    1 &                                                   \\ 
  UX Men B &    1.198 $\pm$    0.007 &   1.274 $\pm$   0.013 &   4.306 $\pm$   0.009 &      EB &      A91 &      F8V&       A91 &    3.781$\pm$     0.007 &    0.287$\pm$     0.029 &      R03 &    1 &                                                   \\ 
  EW Ori A &    1.194 $\pm$    0.014 &   1.141 $\pm$   0.011 &   4.401 $\pm$   0.010 &      EB &      A91 &      G0V&       A91 &    3.776$\pm$     0.007 &    0.170$\pm$     0.030 &      R03 &    1 &                                                   \\ 
  AI Phe A &    1.190 $\pm$    0.006 &   1.762 $\pm$   0.007 &   4.021 $\pm$   0.004 &      EB &      M92 &     K0IV&       A91 &    3.800$\pm$     0.010 &    0.640$\pm$     0.040 &      M92 &    3 &                                                   \\ 
  BH Vir A &    1.165 $\pm$    0.008 &   1.250 $\pm$   0.025 &   4.340 $\pm$   0.020 &      EB &      P97 &  \nodata&       P97 &    3.789$\pm$     0.005 &    0.280$\pm$     0.030 &      R03 &    1 &                                                   \\ 
$\alpha$ Cen A &1.160 $\pm$    0.031 &   \nodata             &   \nodata             &       O &      P00 &      G2V&       P00 &    3.761$\pm$     0.004 &    0.181$\pm$     0.017 &     GD00 &    1 &                                                   \\ 
  EW Ori B &    1.158 $\pm$    0.014 &   1.145 $\pm$   0.011 &   4.384 $\pm$   0.010 &      EB &      P97 &      G5V&       A91 &    3.762$\pm$     0.007 &    0.080$\pm$     0.030 &	  R03 &    1 &                                                   \\ 
  UV Leo B &    1.110 $\pm$    0.100 &   1.216 $\pm$   0.043 &   4.310 $\pm$   0.055 &      EB &      Z03 &  \nodata&       Z03 &    3.759$\pm$     0.004 &    0.140$\pm$     0.045 &      Z03 &    1 &                                                   \\ 
  V432Aur B&    1.060 $\pm$    0.020 &    2.130$\pm$    0.140&     3.810$\pm$    0.060&      EB &    M03   & \nodata &    \nodata&     3.771   $\pm$    0.007 & 0.708 $\pm$ 0.092   &M03  &  3 & \\
  UWLMi A  &    1.060 $\pm$    0.020 &    1.230$\pm$    0.050&     4.280$\pm$    0.030&      EB &    M03   & \nodata &    \nodata&     3.813   $\pm$    0.007 & 0.368 $\pm$ 0.076   &M03  &  1 & \\
V818 Tau A &    1.059 $\pm$    0.006 &   0.900 $\pm$   0.016 &   4.554 $\pm$   0.016 &      EB &     TR02 &      G6V&       G85 &    3.743$\pm$     0.008 &   -0.169$\pm$     0.035 &     TR02 &    1 &                                               Hyades member; CABS    \\ 
  BH Vir B &    1.052 $\pm$    0.006 &   1.140 $\pm$   0.025 &   4.350 $\pm$   0.020 &      EB &      P97 &  \nodata&       P97 &    3.750$\pm$     0.006 &    0.090$\pm$     0.060 &      R03 &    1 &                                                   \\ 
  UWLMi B  &    1.040 $\pm$    0.020 &    1.210$\pm$    0.060&     4.290 $\pm$    0.040&      EB &    M03   & \nodata &    \nodata&     3.813   $\pm$    0.007  & 0.356 $\pm$ 0.080   &M03  &  1 & \\
  CNLyn A  &    1.040 $\pm$    0.020 &    1.800$\pm$    0.210&     3.940 $\pm$    0.100&      EB &    M03   & \nodata &    \nodata&     3.813   $\pm$    0.007  & 0.704 $\pm$ 0.120   &M03  &  3 & \\
  CNLyn B  &    1.040 $\pm$    0.020 &    1.800$\pm$    0.210&     3.940 $\pm$    0.100&      EB &    M03   & \nodata &    \nodata&     3.813   $\pm$    0.007  & 0.704 $\pm$ 0.112   &M03  &  3 & \\
$\chi$ Dra A &  1.030 $\pm$    0.050 &   \nodata             &   \nodata             &       O &      P00 &      F7V&       T87 &    3.742$\pm$     0.015 &    0.258$\pm$     0.047 &      T87 &    3 &                                                   \\ 
       Sun &    1.000 $\pm$    0.000 &   \nodata             &   \nodata             &  \nodata&   \nodata&       G2&    \nodata&    3.761$\pm$     0.001 &    0.000$\pm$     0.007 &      G92 &    1 &                                                   \\ 
  V432Aur A&    0.980 $\pm$    0.020 &    1.390$\pm$    0.080&     4.140$\pm$    0.06&      EB &    M03   & \nodata &    \nodata&     3.785   $\pm$    0.007 & 0.396 $\pm$ 0.088   &M03  &  3 & \\
  UV Psc A &    0.975 $\pm$    0.009 &   1.110 $\pm$   0.020 &   4.335 $\pm$   0.016 &      EB &      P97 &    G4-6V&       S93 &    3.762$\pm$     0.007 &    0.090$\pm$     0.030 &      P97 &    1 &                                             CABS \\ 
$\alpha$ Cen B &0.970 $\pm$    0.030 &   \nodata             &   \nodata             &       O &      P00 &      K1V&       P00 &    3.724$\pm$     0.004 &   -0.300$\pm$     0.011 &     GD00 &    1 &                                                   \\ 
  CG Cyg A &    0.940 $\pm$    0.012 &   0.890 $\pm$   0.013 &   4.512 $\pm$   0.014 &      EB &      P94 &    G9.5V&       S93 &    3.721$\pm$     0.015 &   -0.260$\pm$     0.060 &      P94 &    1 &                                              CABS \\ 
  FL Lyr B &    0.960 $\pm$    0.012 &   0.962 $\pm$   0.028 &   4.454 $\pm$   0.026 &      EB &      A91 &      G8V&       A91 &    3.724$\pm$     0.008 &   -0.180$\pm$     0.040 &      A91 &    1 &                                                   \\ 
$\eta$ Cas A &  0.950 $\pm$    0.080 &   \nodata             &   \nodata             &       O &      F98 &      G3V&       F98 &    3.784$\pm$     0.004 &    0.099$\pm$     0.030 &      F98 &    1 &                                                   \\ 
  RT And B &    0.910 $\pm$    0.020 &   0.900 $\pm$   0.013 &   4.484 $\pm$   0.015 &      EB &      P94 &      K0V&       S93 &    3.675$\pm$     0.010 &   -0.435$\pm$     0.040 &      P94 &    1 &                                              CABS \\ 
  HS Aur A &    0.900 $\pm$    0.019 &   1.004 $\pm$   0.024 &   4.389 $\pm$   0.023 &      EB &      A91 &      G8V&       A91 &    3.728$\pm$     0.006 &   -0.130$\pm$     0.030 &      A91 &    1 &                                                   \\ 
  70 Oph A &    0.900 $\pm$    0.074 &   \nodata             &   \nodata             &       O &      P00 &      K0V&       F98 &    3.726$\pm$     0.002 &   -0.296$\pm$     0.080 &      F98 &    1 &                                                   \\ 
  81 Cnc A &    0.890 $\pm$    0.029 &   \nodata             &   \nodata             &       O &      P00 &      G8V&       M96 &    3.736$\pm$     0.015 &   -0.296$\pm$     0.060 &      M96 &    1 &                                                   \\ 
  HS Aur B &    0.879 $\pm$    0.017 &   0.873 $\pm$   0.024 &   4.500 $\pm$   0.025 &      EB &      A91 &      K0V&       A91 &    3.716$\pm$     0.006 &   -0.300$\pm$     0.030 &      A91 &    1 &                                                   \\ 
$\xi$ Boo A &  0.860 $\pm$    0.070 &   \nodata             &   \nodata             &       O &      F98 &      G8V&       F98 &    3.744$\pm$     0.002 &   -0.272$\pm$     0.030 &      F98 &    1 &                                                   \\ 
  81 Cnc B &    0.850 $\pm$    0.026 &   \nodata             &   \nodata             &       O &      P00 &      G8V&       M96 &    3.736$\pm$     0.015 &   -0.313$\pm$     0.060 &      m96 &    1 &                                                   \\ 
HD195987 A &    0.844 $\pm$    0.018 &   \nodata             &   \nodata             &       O &      To02 &  \nodata&       To02 &    3.716$\pm$     0.008 &   -0.228$\pm$     0.001 &      To02 &    1 & [Fe/H] = -0.5 \\ 
  CG Cyg B &    0.810 $\pm$    0.013 &   0.840 $\pm$   0.014 &   4.505 $\pm$   0.016 &      EB &      P94 &      K3V&       S93 &    3.674$\pm$     0.006 &   -0.510$\pm$     0.030 &      P94 &    1 &                                              CABS \\ 
 HR 6697 B &    0.800 $\pm$    0.055 &   \nodata             &   \nodata             &       O &      P00 &      K3V&      Mc95 &    3.679$\pm$     0.015 &   -0.788$\pm$     0.128 &     Mc95 &    1 &                                                   \\ 
  70 Oph B &    0.780 $\pm$    0.040 &   \nodata             &   \nodata             &       O &      P00 &      K5V&       F98 &    3.638$\pm$     0.015 &   -0.848$\pm$     0.040 &      F98 &    1 &                                                   \\ 
V818 Tau B &    0.760 $\pm$    0.006 &   0.768 $\pm$   0.010 &   4.548 $\pm$   0.011 &      EB &     TR02 &      K6V&       G85 &    3.645$\pm$     0.015 &   -0.775$\pm$     0.062 &     TR02 &    1 &                                               Hyades member; CABS    \\ 
  UV Psc B &    0.760 $\pm$    0.005 &   0.830 $\pm$   0.030 &   4.480 $\pm$   0.031 &      EB &      P97 &   K0-K2V&       S93 &    3.677$\pm$     0.007 &   -0.500$\pm$     0.040 &      P97 &    1 &                                             CABS \\ 
$\chi$ Dra B &  0.730 $\pm$    0.024 &   \nodata             &   \nodata             &       O &      P00 &      K0V&       T87 &    3.719$\pm$     0.030 &   -0.468$\pm$     0.105 &      T87 &    1 &                                                   \\ 
Gl702 B    &    0.713 $\pm$    0.029 &   \nodata             &   \nodata             &       O &      H93 &  \nodata&   \nodata & \textit{3.626$\pm$0.0119} &  -0.805$\pm$    0.05  &      D00   &  1 &                                                   \\
$\xi$ Boo B &  0.700 $\pm$    0.050 &   \nodata             &   \nodata             &       O &      F98 &      K4V&       F98 &    3.638$\pm$     0.015 &   -1.052$\pm$      0.080 &      F98 &    1 &                                                   \\ 
HD195987 B &    0.665 $\pm$    0.008 &   \nodata             &   \nodata             &       O &      To02 &  \nodata&       To02 &    3.623$\pm$     0.021 &   -0.949$\pm$     0.076 &      To02 &    1 & [Fe/H] = -0.5 \\
$\eta$ Cas B &  0.620 $\pm$    0.060 &   \nodata             &   \nodata             &       O &      F98 &      K7V&       F98 &    3.606$\pm$     0.016 &   -1.157$\pm$     0.080 &      F98 &    1 &                                                   \\ 
  YY Gem B &    0.601 $\pm$    0.005 &   0.619 $\pm$   0.006 &   4.632 $\pm$   0.008 &      EB &     TR02 &     dM1e&       S93 &    3.582$\pm$     0.011 &   -1.135$\pm$     0.009 &     TR02 &    1 &                                              CABS \\ 
  YY Gem A &    0.598 $\pm$    0.005 &   0.619 $\pm$   0.006 &   4.632 $\pm$   0.008 &      EB &     TR02 &     dM1e&       S93 &    3.582$\pm$     0.011 &   -1.135$\pm$     0.009 &     TR02 &    1 &                                              CABS \\ 
Gl570 B    &    0.566 $\pm$    0.003 &   \nodata             &   \nodata             &       O &      F99 &  \nodata&    \nodata& \textit{3.548$\pm$0.0056} &  -1.276$\pm$    0.05  &      D00   &  1 &  \\
 CU Cnc Aa &    0.433 $\pm$    0.002 &   0.432 $\pm$   0.005 &   4.804 $\pm$   0.011 &      EB &      R03 &    M3.5V&       R03 &    3.500$\pm$     0.021 &   -1.778$\pm$     0.083 &      R03 &    1 &                                                   \\ 
Gl644 A    &    0.4155$\pm$    0.0057&   \nodata             &   \nodata             &       O &      Se00 & \nodata&    \nodata& \textit{3.524$\pm$0.0036} &  -1.674$\pm$    0.05 &      D00   &  1 &  \\
 CU Cnc Ab &    0.398 $\pm$    0.001 &   0.391 $\pm$   0.009 &   4.854 $\pm$   0.021 &      EB &      R03 &    M3.5V&       R03 &    3.495$\pm$     0.021 &   -1.884$\pm$     0.086 &      R03 &    1 &                                                   \\ 
Gl661 A    &    0.379 $\pm$    0.035 &   \nodata             &   \nodata             &       A &      M98 &  \nodata&    \nodata& \textit{3.509$\pm$0.0028} &  -1.695$\pm$    0.05  &      D00   &  1 &  \\
Gl570 C    &    0.377 $\pm$    0.002 &   \nodata             &   \nodata             &       O &      F99 &  \nodata&    \nodata& \textit{3.519$\pm$0.0083} &  -1.768$\pm$    0.05  &      D00   &  1 &  \\
Gl661 B    &    0.369 $\pm$    0.035 &   \nodata             &   \nodata             &       A &      M98 &  \nodata&    \nodata& \textit{3.526$\pm$0.0039} &  -1.843$\pm$    0.05  &      D00   &  1 &  \\
Gl623 A    &    0.343 $\pm$    0.011 &   \nodata             &   \nodata             &       O &      D00 &  \nodata&    \nodata& \textit{3.531$\pm$0.0032} &  -1.707$\pm$    0.05  &      D00   &  1 &  \\
Gl831 A    &    0.291 $\pm$    0.013 &   \nodata             &   \nodata             &       O &      Se00 & \nodata&    \nodata& \textit{3.486$\pm$0.0022} &  -2.014$\pm$    0.05  &      D00   &  1 &  \\
Gl860 A    &    0.271 $\pm$    0.010 &   \nodata             &   \nodata             &       O &      H99 &  \nodata&    \nodata& \textit{3.507$\pm$0.0023} &  -1.936$\pm$    0.05  &      D00   &  1 &  \\
  CM Dra A &    0.231 $\pm$    0.001 &   0.252 $\pm$   0.002 &   4.998 $\pm$   0.002 &      EB &      M96 &     M4Ve&       S93 &    3.488$\pm$     0.008 &   -2.301$\pm$     0.044 &      V97 &    1 &                                                   \\ 
  CM Dra B &    0.214 $\pm$    0.001 &   0.235 $\pm$   0.002 &   5.025 $\pm$   0.007 &      EB &      M96 &     M4Ve&       S93 &    3.488$\pm$     0.008 &   -2.360$\pm$     0.044 &      V97 &    1 &                                                   \\ 
Gl747 A    &    0.214 $\pm$    0.001 &   \nodata             &   \nodata             &       O &      Se00 & \nodata&    \nodata& \textit{3.508$\pm$0.0026} &  -2.165$\pm$    0.05  &      D00   &  1 &  \\
Gl234 A    &    0.2027 $\pm$   0.0106&   \nodata             &   \nodata             &       O &      D00 &  \nodata&    \nodata& \textit{3.486$\pm$0.0018} &  -2.237$\pm$    0.05  &      D00   &  1 &  \\
Gl747 B    &    0.200 $\pm$    0.001 &   \nodata             &   \nodata             &       O &      Se00 & \nodata&    \nodata& \textit{3.504$\pm$0.0026} &  -2.213$\pm$    0.05  &      D00   &  1 &  \\
Gl860 B    &    0.176 $\pm$    0.007 &   \nodata             &   \nodata             &       O &      H99 &  \nodata&    \nodata& \textit{3.495$\pm$0.0040} &  -2.497$\pm$    0.05  &      D00   &  1 &  \\
Gl831 B    &    0.162 $\pm$    0.007 &   \nodata             &   \nodata             &       O &      Se00 & \nodata&    \nodata& \textit{3.463$\pm$0.0022} &  -2.562$\pm$    0.05  &      D00   &  1 &  \\
Gl473 A    &    0.143 $\pm$    0.011 &   \nodata             &   \nodata             &       A &      T99   &\nodata&    \nodata& \textit{3.455$\pm$0.0021} &  -2.607$\pm$    0.05  &      D00   &  1 &  \\
Gl473 B    &    0.131 $\pm$    0.010 &   \nodata             &   \nodata             &       A &      T99   &\nodata&    \nodata& \textit{3.466$\pm$0.0026} &  -2.745$\pm$    0.05  &      D00   &  1 &  \\
Gl866 B    &    0.1145 $\pm$   0.0012&   \nodata             &   \nodata             &       O &      Se00 & \nodata&    \nodata& \textit{3.453$\pm$0.0021} &  -2.837$\pm$    0.05  &      D00   &  1 &  \\
Gl623 B    &    0.114 $\pm$    0.008 &   \nodata             &   \nodata             &       A &      D00 &  \nodata&    \nodata& \textit{3.453$\pm$0.0042} &  -2.986$\pm$    0.05  &      D00   &  1 &  \\
Gl234 B    &    0.1034 $\pm$   0.0035&   \nodata             &   \nodata             &       O &      Se00 & \nodata&    \nodata& \textit{3.448$\pm$0.0019} &  -2.977$\pm$    0.05  &      D00   &  1 &  \\
 Gl65 A    &    0.102 $\pm$    0.010 &   \nodata             &   \nodata             &       A &      H99 &  \nodata&    \nodata& \textit{3.454$\pm$0.0020} &  -2.754$\pm$    0.05  &      D00   &  1 &  \\
 Gl65 B    &    0.100 $\pm$    0.010 &   \nodata             &   \nodata             &       A &      H99 &  \nodata&    \nodata& \textit{3.453$\pm$0.0022} &  -2.920$\pm$    0.05  &      D00   &  1 &  \\
\hline
\multicolumn{12}{c}{Pre-Main Sequence Stars} \\
\hline
  RS Cha A &    1.858 $\pm$    0.016 &   2.137 $\pm$   0.055 &   4.047 $\pm$   0.023 &      EB &      A91 &       A8&       M00 &    3.883$\pm$     0.010   &   1.144$\pm$     0.044 &      M00 &    2 &                                                   \\ 
  RS Cha B &    1.821 $\pm$    0.018 &   2.338 $\pm$   0.055 &   3.961 $\pm$   0.021 &      EB &      A91 &       A8&       M00 &    3.859$\pm$     0.010   &   1.126$\pm$     0.043 &      M00 &    2 &                                                   \\ 
   MWC 480 &    1.650 $\pm$    0.070 &   \nodata             &   \nodata             &       D &      Si00 &     A2-3&     JJA88 & \textit{3.948$\pm$0.015}  &   1.243$\pm$     0.10  &    \nodata &    2 &                                                 \\ 
  TY CrA B &    1.640 $\pm$    0.010 &   2.080 $\pm$   0.140 &   4.020 $\pm$   0.050 &      EB &      C98 &  \nodata&       C98 &    3.690$\pm$     0.035   &   0.380$\pm$     0.145 &      C98 &    2 &                             \\ 
045251+3016 A & 1.450 $\pm$    0.190 &   \nodata             &   \nodata             &       O &     St01 &       K5&      St01 & \textit{3.643$\pm$0.015}  &  -0.167$\pm$     0.053 &     St01 &    2 &                                                   \\ 
  AK Sco A   &   1.350 $\pm$0.070       &    1.590 $\pm$    0.350 &   \nodata             &      SB &      A03 &      F5 &       A89 &   3.813$\pm$  0.007   &    0.607$\pm$   0.050    &      A03 &    2 &  \\
  AK Sco B   &   1.350 $\pm$0.070       &    1.590 $\pm$    0.350 &   \nodata             &      SB &      A03 &      F5 &       A89 &   3.813$\pm$  0.007   &    0.607$\pm$   0.050    &      A03 &    2 &  \\
    BP Tau &    1.320 $\pm$    0.200 &   \nodata             &   \nodata             &       D &      Du03 &       K7&    B90/H95 &    3.608$\pm$0.012        &  -0.780$\pm$     0.10  &      JVK99 &    2 &                                                  \\ 
0529.4+0041 A & 1.250 $\pm$    0.050 &   1.700 $\pm$   0.200 &   4.070 $\pm$   0.100 &      EB &      C00 &    K1-K2&       C00 &    3.701$\pm$     0.009   &   0.243$\pm$     0.037 &      C00 &    2 &                                                   \\ 
  EK Cep B &    1.124 $\pm$    0.012 &   1.320 $\pm$   0.015 &   4.250 $\pm$   0.015 &      EB &      P87 &  \nodata&    \nodata&    3.755$\pm$     0.015   &   0.190$\pm$     0.070 &      P87 &    2 &                                                   \\
 UZ Tau Aa &    1.016 $\pm$    0.065 &   \nodata             &   \nodata             &     DSB &      Si00 &       M1&       P02 & \textit{3.557$\pm$0.015}  &  -0.201$\pm$     0.124 &      P02 &    2 &                                                   \\ 
0529.4+0041 B & 0.910 $\pm$    0.050 &   1.200 $\pm$   0.200 &   4.240 $\pm$   0.150 &      EB &      C00 &    K7-M0&       C00 &    3.604$\pm$     0.022   &  -0.469$\pm$     0.192 &      C00 &    2 &                                                   \\ 
   LkCa 15 &    0.970 $\pm$    0.030 &   \nodata             &   \nodata             &       D &      Si00 &       K5&       H86 & \textit{3.643$\pm$0.015}  &  -0.165$\pm$     0.10  &   \nodata&    2 &                                                \\ 
    GM Aur &    0.840 $\pm$    0.050 &   \nodata             &   \nodata             &       D &      Si00 &       K7&    B90/H95 & \textit{3.602$\pm$0.015}  &   0.598$\pm$     0.10  &   \nodata&    2 &                                                \\ 
045251+3016 B & 0.810 $\pm$    0.090 &   \nodata             &   \nodata             &       O &     St01 &  \nodata&    \nodata& \textit{3.535$\pm$0.015}  &  -0.830$\pm$     0.086 &     St01 &    2 &                                                   \\ 
    DL Tau &    0.720 $\pm$    0.110 &   \nodata             &   \nodata             &       D &      Si00 &    K7-M0&    B90/H95 & \textit{3.591$\pm$0.015}  &   0.005$\pm$     0.10  &   \nodata&    2 &                                                \\ 
    DM Tau &    0.550 $\pm$    0.030 &   \nodata             &   \nodata             &       D &      Si00 &       M1&       V93 & \textit{3.557$\pm$0.015}  &  -0.532$\pm$     0.10  &   \nodata&    2 &                                                \\ 
    CY Tau &    0.550 $\pm$    0.330 &   \nodata             &   \nodata             &       D &      Si00 &       M2&      SS94 & \textit{3.535$\pm$0.015}  &  -0.491$\pm$     0.10  &   \nodata&    2 &                                                \\ 
 UZ Tau Ab &    0.294 $\pm$    0.027 &   \nodata             &   \nodata             &     DSB &      Si00 &       M4&       P02 & \textit{3.491$\pm$0.015}  &  -0.553$\pm$     0.124 &      p02 &    2 &        \\ 
\hline
\multicolumn{12}{c}{Pre-Main Sequence Composite Systems} \\
\hline
 GG Tau Aa &    (1.28 $\pm$    0.07)$\times C$ &   \nodata   &   \nodata             &       D &      Si01 &       M0&      HK03 & \textit{3.580$\pm$0.015} &   -0.106$\pm$     0.10  &    \nodata  &    2 &                                                \\ 
 GG Tau Ab &    (1.28 $\pm$    0.07)$\times (1-C)$ &\nodata  &   \nodata             &       D &      Si01 &       M2&      HK03 & \textit{3.535$\pm$0.015} &   -0.338$\pm$     0.10  &\nodata  &    2 &                                                \\ 
  DF Tau A &    (0.90 $\pm$    0.60)$\times C$ &   \nodata   &   \nodata             &       A &      S03 &       M2&      HK03 & \textit{3.535$\pm$0.015} &   -0.255$\pm$     0.10  &    \nodata  &    2 &                                                \\ 
  DF Tau B &    (0.90 $\pm$    0.60)$\times (1-C)$ &\nodata  &   \nodata             &       A &      S03 &     M2.5&      HK03 & \textit{3.524$\pm$0.015} &   -0.162$\pm$     0.10  &\nodata  &    2 &                                                \\ 
  FS Tau A &    (0.78 $\pm$    0.25)$\times C$ &   \nodata   &   \nodata             &       A &      Ta02 &       M0&      HK03 & \textit{3.580$\pm$0.015} &   -1.293$\pm$     0.10  &    \nodata  &    2 &                                                \\ 
  FS Tau B &    (0.78 $\pm$    0.25)$\times (1-C)$ &\nodata  &   \nodata             &       A &      Ta02 &     M3.5&      HK03 & \textit{3.502$\pm$0.015} &   -1.552$\pm$     0.10  &\nodata  &    2 &                                                \\ 
  FO Tau A &    (0.77 $\pm$    0.25)$\times C$ &   \nodata   &   \nodata             &       A &      Ta02 &     M3.5&      HK03 & \textit{3.502$\pm$0.015} &   -0.581$\pm$     0.10  &    \nodata &    2 &                                                \\ 
  FO Tau B &    (0.77 $\pm$    0.25)$\times (1-C)$ &\nodata  &   \nodata             &       A &      Ta02 &     M3.5&      HK03 & \textit{3.502$\pm$0.015} &   -0.609$\pm$     0.10  & \nodata &    2 &                                                \\ 
\enddata
\tablenotetext{a}{\,Method used to determine the dynamical mass:
EB = eclipsing binary system,
O = astrometric + radial velocity orbit,
A = astrometric orbit + distance estimate,
D =  disk kinematics,
DSB = disk kinematics + doubled-lined spectroscopic binary}

\tablenotetext{b}{\,Main-sequence temperatures in italics are determined from colors; pre-main sequence 
temperatures in italics are determined from a spectral type.}

\tablenotetext{c}{\,Evolutionary status code:  
1 = main sequence;
2 =  pre-main sequence;
3 = post-main sequence / evolved.}

\tablerefs{
References: A91 = Anderson (1991);
A03 = Alencar et al. (2003);
A89 = Andersen et al. (1989);
B90 = Basri \& Batalha (1990); 
C98 = Casey et al. (1998); 
C00 = Covino et al. (2000); 
D00 = Delfosse, X. et al. (2000);
Dr03 = Drummond et al. (2003); 
Du03 = Dutrey et al. (2003); 
F98 = Fernandes et al. (1998);
F99 = Forveille et al. (1999); 
G85 = Griffin (1985);
Gl69 = Gliese (1969); 
GD00 = Guenther \& Demarque (2000); 
H95 = Hartigan et al. (1995);
H86 = Herbig et al. (1986);
H93 = Henry et al. (1993);
H99 = Henry et al. (1999);
HE84 = Hill \& Ebbighausen (1984);
HK03 = Hartigan \& Kenyon (2003);
JJA88 =  Jaschek et al. (1988);
L96 = Latham (1996);
L97a = Lacy (1997a);
L97b = Lacy (1997b);
L97c = Lacy (1997c);
L97d = Lacy (1997d);
Mc95 = McAlister et al. (1995);
M98 = Martin et al. (1998); 
M01 = Munari (2001);
M03 = Marrese et al. (2003);
M92 = Milone et al. (1992);
P87 = Popper (1987);
P02 = Prato et al. (2002);
P94 = Popper (1994);
R99 = Ribas et al. (1999);
Se00 = S\'\ e gransan et al. (2000);
Si00 = Simon et al. (2001);
St01 = Steffen et al. (2001);
SS94 = Stuwe \& Schulz (1994);
S03 = Schaefer et al. (2003);
S93 = Strassmeier et al. (1993);
Ta02 = Tamazian et al(2002);
T87 = Tomkin et al. (1987);
T97a = Torres et al. (1997a);
T97b = Torres et al. (1997b);
T99 = Torres et al. (1999); 
TR02 = Torres \& Ribas (2002);
To02 = Torres et al. (2002);
V93 = Valenti et al. (1993);
Z03 = Zwitter et al. (2003)
}

\end{deluxetable}

\end{document}